\documentclass[preprints,article,accept,moreauthors,pdftex]{Definitions/mdpi}

\firstpage{1} 
\pubvolume{xx}
\issuenum{1}
\articlenumber{5}
\pubyear{2019}
\copyrightyear{2019}
\history{}
\usepackage{todonotes}
\usepackage{xspace}
\usepackage{amsmath}
\usepackage{amssymb}
\usepackage{footmisc}
\usepackage{subcaption}
\usepackage{placeins}
\usepackage{booktabs}

\usepackage{tabularx}
\usepackage{multirow}
\usepackage{rotating}
\usepackage{colortbl}
\usepackage[table]{xcolor}
\usepackage{cancel}

\Title{Faster and Better Nested Dissection Orders for Customizable Contraction Hierarchies}
\Author{Lars Gottesbüren $^{1}$, Michael Hamann $^{2}$, Tim Niklas Uhl $^{3}$ and Dorothea Wagner $^{4}$}

\AuthorNames{Lars Gottesbüren, Michael Hamann, Tim Niklas Uhl and Dorothea Wagner}

\address{%
  $^{1}$ \quad Institute of Theoretical Informatics, Karlsruhe Institute of Technology, Am Fasanengarten 5, 76131 Karlsruhe, Germany; lars.gottesbueren@kit.edu\\
  $^{2}$ \quad Institute of Theoretical Informatics, Karlsruhe Institute of Technology, Am Fasanengarten 5, 76131 Karlsruhe, Germany; michael.hamann@kit.edu\\
  $^{3}$ \quad Institute of Theoretical Informatics, Karlsruhe Institute of Technology, Am Fasanengarten 5, 76131 Karlsruhe, Germany; niklas.uhl@online.de\\
  $^{4}$ \quad Institute of Theoretical Informatics, Karlsruhe Institute of Technology, Am Fasanengarten 5, 76131 Karlsruhe, Germany; dorothea.wagner@kit.edu\\
}

\abstract{Graph partitioning has many applications. We consider the acceleration of shortest path queries in road networks using Customizable Contraction Hierarchies (CCH).
  It is based on computing a nested dissection order by recursively dividing the road network into parts.
  Recently, with FlowCutter and Inertial Flow, two flow-based graph bipartitioning algorithms have been proposed for road networks.
  While FlowCutter achieves high-quality results and thus fast query times, it is rather slow.
  Inertial Flow is particularly fast due to the use of geographical information while still achieving decent query times.
  We combine the techniques of both algorithms to achieve more than six times faster preprocessing times than FlowCutter and even faster queries on the Europe road network.
  We show that using 16 cores of a shared-memory machine, this preprocessing needs four minutes.}

%keyword 1; keyword 2; keyword 3 (list three to ten pertinent keywords specific to the article, yet reasonably common within the subject discipline.)}

\newcommand{\ie}{i.\,e.\xspace}
\newcommand{\eg}{e.\,g.\xspace}

\DeclareMathOperator*{\cut}{cut}
\renewcommand{\epsilon}{\varepsilon}

\newlength{\mycolwidth}
\newlength{\mysmallcolwidth}

\begin{document}
%%%%%%%%%%%%%%%%%%%%%%%%%%%%%%%%%%%%%%%%%%

\section{Introduction}

The goal of graph partitioning is to divide a graph into a given number of roughly equally sized parts by removing a small number of edges or nodes.
Graph partitioning has many practical applications such as accelerating matrix multiplication, dividing compute workloads, image processing, VLSI design and, the focus of this work, accelerating shortest path computations in road networks.
For an overview of the state of the art in graph partitioning we refer the reader to a survey article~\cite{bmsss-ragp-16}.

Modern speedup techniques for shortest path computation in road networks usually achieve fast queries by an expensive preprocessing phase, which builds a metric-independent index datastructure and a customization phase, which incorporates the metric (e.g. travel time, walking distance, and real-time traffic information) into the index.
The query phase uses the index to answer queries very quickly.
For such three-phase approaches, the preprocessing phase typically includes a graph partitioning step, \eg a hierarchy of nested partitions or a nested dissection~\cite{g-ndrfe-73} order.

Contraction Hierarchies simulate contracting all nodes in a given order and insert shortcut arcs between the neighbors of the contracted node.
These represent paths via the contracted nodes.
Shortest $s-t$ path queries are answered by \eg a bidirectional Dijkstra search~\cite{d-ntpcg-59} from $s$ and $t$, which only considers shortcut and original arcs to higher-ranked nodes.
Thus, nodes which lie on many shortest paths should be ranked highly in the order.
Customizable Contraction Hierarchies~\cite{dsw-cch-15} use contraction orders computed via recursive balanced node separators (nested dissection) in order to achieve a logarithmic search space depth with few added shortcuts.
Node separators are considered to lie on many shortest paths, as any path between the components crosses the separator.
The weights of the contraction hierarchy can then be quickly customized to any metric, allowing to, e.g., incorporate real-time traffic information.
The running time needed for the customization and the shortest path queries depends on the quality of the calculated order.
Previously proposed partitioning tools for computing separators in road networks include FlowCutter~\cite{hs-gbpo-18}, Inertial Flow~\cite{ss-obsrn-15}, KaHiP~\cite{ss-tlagh-13} and Metis~\cite{kk-mspig-99}, PUNCH~\cite{dgrw-gpnc-11} and Buffoon~\cite{ss-degp-12}.
KaHiP and Metis are general-purpose graph partitioning tools.
PUNCH and Buffoon are special-purpose partitioners, which aim to use geographical features of road networks such as rivers or mountains.
Rivers and mountains form very small cuts and were dubbed \emph{natural cuts} in~\cite{dgrw-gpnc-11}.
PUNCH identifies and deletes natural cuts, then contracts the remaining components, and subsequently runs a variety of highly randomized local search algorithms.
Buffoon incorporates the idea of natural cuts into KaHiP, running its evolutionary multilevel partitioner instead of the flat local searches of PUNCH.
In~\cite{hs-gbpo-18} it was shown that FlowCutter is also able to identify and leverage natural cuts.
Inertial Flow is another special-purpose partitioner that is even based on using the geographic embedding of the road network.

We combine the idea of Inertial Flow to use geographic coordinates with the incremental cut computations of FlowCutter.
This allows us to compute a series of cuts with suitable balances much faster than FlowCutter while still achieving high quality.
In an extensive experimental evaluation, we compare our new algorithm \emph{InertialFlowCutter} to the state-of-the-art.
FlowCutter is the previously best method for computing CCH orders.
InertialFlowCutter computes slightly better CCH orders than FlowCutter and is a factor of 5.7 and 6.6 faster on the road networks of the USA and Europe, respectively -- our two most relevant instances.
Using 16 cores of a shared-memory machine we can compute CCH orders for these instances in four minutes.

In Section~\ref{sec:materials_and_methods} we briefly present the existing Inertial Flow and FlowCutter algorithms and describe how we combined them.
In Section~\ref{sec:results} we describe the setup and results of our experimental study.
We conclude with a discussion of our results and future research directions in Section~\ref{sec:discussion}.

This paper recreates the experiments from~\cite{hs-gbpo-18} and uses a lot of the same setup.
Therefore, there is substantial content overlap.
To keep this paper self-contained, we repeat the parts we use.
Our contributions are the InertialFlowCutter algorithm, an improved Inertial Flow implementation and a reproduction of the experiments from~\cite{hs-gbpo-18}, including InertialFlowCutter and a newer KaHiP version.

%%%%%%%%%%%%%%%%%%%%%%%%%%%%%%%%%%%%%%%%%%

\section{Materials and Methods}\label{sec:materials_and_methods}

After introducing preliminaries, we describe the existing biparitioning algorithms FlowCutter and Inertial Flow on a high level, before discussing how to combine them into our new algorithm InertialFlowCutter.
We refer the interested reader to~\cite{hs-gbpo-18} for implementation details and a more in-depth discussion of the FlowCutter algorithm.
Then we discuss our application Customizable Contraction Hierarchies (CCH), what makes a good CCH order, and how we use recursive bisection to compute them.

\subsection{Preliminaries}

An undirected graph $G=(V,E)$ consists of a set of \emph{nodes} $V$ and a set of \emph{edges} $E \subseteq {{V}\choose{2}}$.
A directed graph $G=(V,A)$ has directed \emph{arcs} $A \subseteq V \times V$ instead of undirected edges.
It is symmetric iff for every arc $(x,y) \in A$ the reverse arc $(y,x) \in A$ exists.
For ease of notation, we do not distinguish between undirected and symmetric graphs in this paper, and we use them interchangeably, whichever better suits the description.
Let $n := |V|$ denote the number of nodes and let $m := |E|$ denote the number of edges of an undirected graph.
All graphs in this paper contain neither self-loops $(x,x)$ nor multi-edges.
$H=(V',A')$ is a subgraph of $G$, iff $V' \subseteq V$ and $A' \subseteq A$.
The subgraph induced by a node set $U \subseteq V$ is defined as $G[U] := (U, \{(u,v) \in A \cap (U \times U)\})$, the graph with nodes $U$ and all arcs/edges of $G$ with endpoints in $U$.
The degree $\deg(x) = |\{(x,y) \in A\}|$ is the number of \emph{outgoing} arcs of $x$.
A path is a sequence of edges such that consecutive edges overlap in a vertex.
A graph is called $k$-connected, iff there are $k$ node-disjoint paths between every pair of nodes.
The $k$-connected components of a graph are the node-induced subgraphs, which are inclusion-maximal regarding $k$-connectivity.
$1$-connected components are called connected components, $2$-connected components are called biconnected components.

\paragraph{Separators and Cuts}
Let $V_1, V_2 \subset V$ be a \emph{bipartition} of $V = V_1 \cup V_2$ into two non-empty disjoint sets, called \emph{blocks}.
The \emph{cut} induced by $(V_1,V_2)$ is the set of edges $\cut(V_1,V_2):= \{ (v_1,v_2) \subseteq E \cap (V_1 \times V_2) \}$ between $V_1$ and $V_2$.
The cut size is $|\cut(V_1,V_2)|$.
We often use the terms cut and bipartition interchangeably.
Sometimes we say a bipartition is induced by a set of cut edges.

A node separator partition is a partition of $V = Q \cup V_1 \cup V_2$ into three disjoint sets $(Q,V_1,V_2)$ such that there is no edge between $V_1$ and $V_2$.
We call $Q$ the separator and $V_1,V_2$ the blocks or components of the separator.
$|Q|$ is the separator size.

For an $\epsilon \in [0,1]$, a cut or separator is $\epsilon$-balanced if $\max(|V_1|,|V_2|) \leq \lceil (1+\epsilon)n/2 \rceil$.
We often call $\epsilon$ the \emph{imbalance}, as larger values correspond to less balanced cuts.
The \emph{balanced graph bipartitioning} [\emph{balanced node separator}] problem is to find an $\epsilon$-balanced cut [separator] of minimum size.

Let $S,T \subset V$ be two fixed, disjoint, non-empty subsets of $V$.
An edge cut [node separator] is an $S$-$T$ edge cut [node separator] if $S \subseteq V_1$ and $T \subseteq V_2$.

\paragraph{Maximum Flows}

A flow network $\mathcal{N}=(V,A,S,T,c)$ is a simple symmetric directed graph $(V,A)$ with two disjoint non-empty \emph{terminal} node sets $S,T\subsetneq V$, also called the source and target node set, as well as a capacity function $c : A \mapsto \mathbb{R}_{\geq 0}$.
A flow in $\mathcal{N}$ is a function $f:A \mapsto \mathbb{R}$ subject to the \emph{capacity constraint} $f(a) \leq c(a)$ for all arcs $a$, \emph{flow conservation} ${\sum_{(u,v)\in A} f((u,v)) = 0}$ for all non-terminal nodes $v$ and \emph{skew symmetry} ${f((u,v))=-f((v,u))}$ for all arcs~$(u,v)$.
In this paper we consider only unit flows and unit capacities, \ie $f : A \mapsto \{-1, 0,1\}$, $c : A \mapsto \{0, 1\}$.
The \emph{value} of a flow ${|f| := \sum_{s \in S, (s,u)\in A} f((s,u))}$ is the amount of flow leaving $S$.
The \emph{residual capacity} $r_f(a) := c(a) - f(a)$ is the additional amount of flow that can pass through $a$ without violating the capacity constraint.
The residual network with respect to $f$ is the directed graph $\mathcal{N}_f = (V,A_f)$ where $A_f := \{a \in A | r_f(a) > 0\}$.
An \emph{augmenting path} is an $S$-$T$ path in $\mathcal{N}_f$.
A node $v$ is called \emph{source-reachable} if there is a path from $S$ to $v$ in $\mathcal{N}_f$.
We denote the set of source-reachable nodes by $S_r$, and define the set of \emph{target-reachable} nodes $T_r$ analogously.
The flow $f$ is a \emph{maximum flow} if $|f|$ is maximal among all possible flows in $\mathcal{N}$.
This is the case iff there is no augmenting path in $\mathcal{N}_f$.
The well-known max-flow-min-cut theorem~\cite{ff-mftn-56} states that the value of a maximum flow equals the capacity of a minimum $S$-$T$ edge cut.
$(S_r, V\setminus S_r)$ is the source-side cut and $(V\setminus T_r, T_r)$ is the target-side cut of a maximum flow.

\subsection{FlowCutter}\label{sec:flowcutter}
FlowCutter is an algorithm for the balanced graph bipartitioning problem.
The idea of its \emph{core algorithm} is to solve a sequence of incremental max flow problems, which induce cuts with monotonically increasing cut size and balance, until the latest cut induces an $\epsilon$-balanced bipartition.
The flow problems are incremental in the sense that the terminal nodes $S,T$ of the previous flow problem are subsets of the terminals in the next flow problem.
This nesting allows us to reuse the flow computed in previous iterations.

Given starting terminal nodes $s,t$, we set $S:=\{s\}, T:=\{t\}$ and compute a maximum $S$-$T$ flow.
Then we transform the $S$-reachable nodes $S_r$ to sources, if $S_r \leq T_r$, or $T_r$ to targets otherwise.
Assume $S_r \leq T_r$ without loss of generality.
Now $S$ induces a minimum $S$-$T$ cut $C_S$.
If $C_S$ is $\epsilon$-balanced, we terminate.
Otherwise we transform one additional node, called \emph{piercing node}, to a source.
The piercing node is chosen from the nodes incident to the cut $C_S$ and not in $S$.
This step is called \emph{piercing} the cut $C_S$.
It ensures we find a different cut in the next iteration.
Subsequently, we augment the previous flow to a maximum flow which considers the new source node.
These steps are repeated until the latest cut induces an $\epsilon$-balanced bipartition.

A significant detail of the piercing step is that piercing nodes which are not reachable from the opposite side are preferred.
Choosing such nodes for piercing does not create augmenting paths.
Thus the cut size does not increase in the next iteration.
This is called the \emph{avoid-augmenting-paths} heuristic.
A secondary \emph{distance-based} piercing heuristic is used to break ties, when the avoid-augmenting-paths heuristic gives multiple choices.
It chooses the node $p$ which minimizes $\operatorname{dist}(p,t) - \operatorname{dist}(s,p)$, where $\operatorname{dist}$ is the hop distance, precomputed via Breadth-First-Search from $s$ and $t$.
Roughly speaking, this attempts to prevent the cut sides from meeting before perfect balance.
It also has a geometric interpretation, which is explained in~\cite{hs-gbpo-18}.

We choose the starting terminal nodes $s$ and $t$ uniformly at random.
Experiments~\cite{hs-gbpo-18} indicate that 20 terminal pairs are sufficient to obtain high quality partitions of road networks.

For computing maximum flows, we use the basic Ford-Fulkerson algorithm~\cite{ff-mftn-56}, with Pseudo-Depth-First-Search for finding augmenting paths.
Pseudo-Depth-First-Search directly marks all adjacent nodes as visited when processing a node.
It can be implemented like Breadth-First-Search by using a stack instead of a queue.

A major advantage of FlowCutter over other partitioning tools is the fact that it computes multiple cuts, which form a Pareto cutset after filtering dominated cuts.
By this, we mean for every pair of cuts $C_1, C_2$ in the Pareto cutset, the cut $C_1$ either has fewer edges or has better balance than $C_2$.
This means that we do not need to determine the maximum imbalance a priori, but we can select a good trade-off between cut size and imbalance from the Pareto cutset.

\subsection{Inertial Flow}\label{sec:inertialflow}

Given a line $l \in \mathbb{R}^2$, Inertial Flow orthogonally projects the nodes onto $l$, according to their geographical coordinates.
The nodes are sorted by order of appearance on $l$.
For a parameter $\alpha \in [0, 0.5]$ the first $\lfloor \alpha \cdot n \rfloor$ nodes are chosen as $S$.
Analogously, the last $\lfloor \alpha \cdot n \rfloor$ nodes are chosen as $T$.
In the next step, a maximum $S$-$T$ flow is computed from which a minimum $S$-$T$ cut is derived.
Instead of line, we use the term \emph{direction}.
In~\cite{ss-obsrn-15}, $\alpha=0.2$ and four directions are used: West-East, South-North, Southwest-Northeast and Southeast-Northwest.
This simple approach works surprisingly well for road networks.

\subsection{Combining Inertial Flow and FlowCutter into InertialFlowCutter}\label{sec:inertialflowcutter}

One drawback of Inertial Flow is the restriction to one value of $\alpha$.
We enhance FlowCutter by initializing $S$ and $T$ in the same way as Inertial Flow, however with a smaller parameter $\alpha$ than proposed for Inertial Flow.
Additionally, we pierce cuts with multiple nodes from the Inertial Flow order at once.
We call this \emph{bulk piercing}.
This way, we enumerate multiple Inertial Flow cuts simultaneously, without having to restart the flow computations.
Furthermore, we can skip some of the first, highly imbalanced cuts of FlowCutter that are irrelevant for our application.

We introduce three additional parameters $\gamma_a, \gamma_o \in (0, 0.5]$ and $\delta \in (0,1)$ to formalize bulk piercing.
Let $L$ be a permutation of the nodes, ordered according to a direction.
For the source side, we use bulk piercing as long as $S$ contains at most $\gamma_a \cdot n$ nodes.
Further, we limit ourselves to piercing the first $\gamma_o \cdot n$ nodes of $L$.
The parameter $\delta$ influences the step size.
The idea is to decrease the step size as our cut becomes more balanced.
When we decide to apply bulk piercing, we settle the next $\delta (\frac{1-\delta}{2}n - |S|)$ nodes to $S$, when piercing the source side.
To enforce the limit set by $\gamma_o$, we pierce fewer nodes if necessary.
For the target side, we apply this analogously starting from the end of the order.
If bulk piercing cannot be applied, we revert to the standard FlowCutter method of selecting single piercing nodes incident to the cut.
Additionally, we always prioritize the avoid-augmenting-paths heuristic over bulk piercing.

In our experiments, we conduct a parameter study which yields $\alpha = 0.05, \gamma_a = 0.4, \gamma_o = 0.25$ and $\delta = 0.05$ as reasonable choices.

\subsection{Running Multiple InertialFlowCutter Instances}

To improve solution quality, we run $q \in \mathbb{N}$ instances of InertialFlowCutter with different directions.
An instance is called a \emph{cutter}.
We use the directions $(\cos(\varphi), \sin(\varphi))$ for $\varphi = \frac{k \pi}{q}$ and $k \in [0,\dots,q-1]$.
To include the directions proposed in~\cite{ss-obsrn-15}, $q$ should be a multiple of $4$.
To improve running time, we run cutters simultaneously in an \emph{interleaved} fashion as already proposed in~\cite{hs-gbpo-18}.
We always schedule the cutters with the currently smallest flow value to either push one additional unit of flow or derive a cut.
For the latter, we improve the balance by piercing the cut as long as this does not create an augmenting path.
One stand-alone cutter runs in $\mathcal{O}(cm)$, where $c$ is the size of the largest output cut.
Roughly speaking, this stems from performing one graph traversal,~\eg Pseudo-DFS, per unit of flow.
The exact details can be found in~\cite{hs-gbpo-18}.
Flow-based execution interleaving ensures that no cutter performs more flow augmentations than the other cutters.
Thus, the running time for $q$ cutters is $\mathcal{O}(qcm)$, where $c$ is the size of the largest found cut among all cutters.
We specifically avoid computing some cuts that the stand-alone cutters would find.
Consider the simple example with $q=2$, where the second cutter immediately finds a perfectly balanced cut with cut size $c$ but the first cutter only finds one cut with cut size $C \gg c$.
If the first cutter runs until a cut is found, we invested $Cm$ work, but should only have invested $cm$.

In the case of InertialFlowCutter it is actually important to employ flow-based interleaving and not just run a cutter until the next cut is found, as after a bulk piercing step the next cut might be significantly larger.
For road networks and FlowCutter, this difference is insignificant in practice, as the cut increases by just one, most of the time.

\subsection{Customizable Contraction Hierarchies}\label{sec:cch}

A Customizable Contraction Hierarchy (CCH) is an index data structure which allows fast shortest path queries and fast adaptation to new metrics in road networks.
It consists of three phases: a \emph{preprocessing} phase, which only uses the network topology, a faster \emph{customization} phase, which adapts the index to the weights of the edges, and a \emph{query} phase which quickly answers shortest path queries.
The preprocessing phase simulates contracting all nodes in a given order and inserts shortcut arcs between all neighbors of the contracted node.
The customization phase assigns correct weights to shortcuts by processing all arcs $(u,v)$ in the order ascending by rank of $u$, \ie{}, the position of $u$ in the order.
To process an arc $(u,v)$, it enumerates all triangles $\langle u,w,v \rangle$ where $w$ has lower rank than $u$ and $v$, and updates the weight of $(u,v)$ if the path $(u,w,v)$ is shorter.
There are two different algorithms for $s$-$t$ queries.
The first, basic query algorithm performs bidirectional Dijkstra search from $s$ and $t$ and relaxes only arcs to higher-ranked nodes.
The second query algorithm uses the \emph{elimination tree} of a CCH to avoid priority queues, which are typically a bottleneck.
In the elimination tree, the parent of a node is its lowest-ranked upward neighbor.
The ancestors of a node $v$ are exactly the nodes in the upward search space of $v$ in the basic query~\cite{bcrw-s-16}.
For the $s$-$t$ query, the outgoing arcs of all nodes on the path from $s$ to the root and all incoming arcs of all nodes on the path from $t$ to the root are relaxed.
The node $z$ minimizing the distance from $s$ to $z$ plus the distance from $z$ to $t$ determines the distance between $s$ and $t$.

The query complexity is linear in the number of arcs incident to nodes on the paths from $s$ and $t$ to the root.
Similarly, the customization running time depends on the number of triangles in the CCH.
Fewer shortcuts result in less memory consumption and faster queries.
We aim to minimize these metrics by computing high quality contraction orders.

\subsection{Nested Dissection Orders For Road Networks}\label{sec:nested_dissection}

The framework to compute contraction orders is the same as for FlowCutter in~\cite{hs-gbpo-18}.
For self-containedness we repeat it here.
We only exchange the partitioning algorithm.

\paragraph{Recursive Bisection}
We compute contraction orders via recursive bisection, using node separators instead of edge cuts.
This method is also called nested dissection~\cite{g-ndrfe-73}.
Let $(Q,V_1,V_2)$ be a node separator partition.
Then we recursively compute orders for $G[V_1]$ and $G[V_2]$ and return the order of $G[V_1]$ followed by the order of $G[V_2]$ followed by $Q$.
$Q$ can be in an arbitrary order.
We opt for the input order.
Recursion stops once the graphs are trees or cliques.
For cliques, any order is optimal.
For trees, we use an algorithm to compute an order with minimal elimination tree depth in linear time~\cite{irv-o-88,s-o-89}.

\paragraph{Separators}
InertialFlowCutter computes edge cuts.
We use a standard construction~\cite{amo-nf-93} to model node capacities as edge capacities in flow networks -- which corresponds to node separators as edge cuts.
It expands the undirected input graph $G=(V,E)$ into a directed graph $G'=(V',A')$.
For every node $v \in V$, there is an \emph{in-node} $v_i$ and an \emph{out-node} $v_o$ in $V'$, joined by a directed arc $(v_i, v_o)$, called the  \emph{bridge arc} of $v$.
Further, for every edge $\{u,v\} \in E$ there are two directed \emph{external} arcs $(u_o, v_i)$ and $(v_o, u_i) \in A'$.
Since we restrict ourselves to unit capacity flow networks, we cannot use infinite capacity for external arcs and our cuts contain both bridge arcs and external arcs.
Bridge arcs directly correspond to a node in the separator.
From the external cut arcs, the incident node on the larger side of the cut is included in the separator.

\paragraph{Choosing Cuts from the Pareto Cutset}
InertialFlowCutter yields a sequence of non-dominated cuts with monotonically increasing cut size and balance, whereas standard partitioners yield a single cut for some prespecified imbalance.
We need to choose one cut, to recurse on the sides of the corresponding separator.
The \emph{expansion} of a cut is its cut size divided by the number of nodes on the smaller side.
This gives a certain trade-off between cut size and balance.
We choose the cut with minimum expansion and $\epsilon < 0.6$, \ie at least 20\% of the nodes on the smaller side.
While this approach is certainly not optimal, it works well enough.
It is not clear how to choose the optimum cut without considering the whole hierarchy of cuts in deeper levels of recursion.

\paragraph{Special Preprocessing}
Road networks contain many nodes of degree 1 or 2.
The graph size can be drastically reduced by eliminating them in a preprocessing step that is performed only once.
First we compute the largest biconnected component $B$ and remove all edges between $B$ and the rest of the graph $G$.
The remaining graph usually consists of a large $B$ and many tiny, often tree-like components.
We compute orders for the components separately and concatenate them in an arbitrary order.
The order for $B$ is placed after the orders of the smaller components.

A degree-2-chain is a path $(x,y_1,\dots,y_k,z)$ where all $\deg(y_i) = 2$ but $\deg(x) > 2$ and $\deg(z) \neq 2$.
We divide the nodes into two graphs $G_{\geq 3}$ and $G_{\leq 2}$ with degrees at least $3$ and at most $2$, by computing all degree-2-chains in linear time and splitting along them.
If $\deg(z) > 2$, we insert an edge between $x$ and $z$ since $z$ is in $G_{\geq 3}$.
We compute contraction orders for the connected components of $G_{\leq 2}$ separately and concatenate them in an arbitrary order.
Since these are paths, we can use the algorithm for trees.
The order for $G_{\geq 3}$ is placed after the one for $G_{\leq 2}$.

\subsection{Parallelization}\label{sec:parallelization}

Recursive bisection is straightforward to parallelize by computing orders on the separated blocks independently, using task-based parallelism.
This only employs parallelism after the first separators have been found.
Therefore, we additionally parallelize InertialFlowCutter.
The implementation of FlowCutter~\cite{flowcutter-github} contains a simple parallelization that lets all cutters with minimum cut progress to the next cut in a parallel for loop.

Recall that we interleave cutter execution based on flow, not on cuts.
Waiting after every flow unit incurs too much idle time in practice, as different flow augmentations take different amounts of time.
We employ a more sophisticated parallelization, using task-based parallelism.
For $q$ cutters, we create $q$ tasks and leave it up to the non-preemptive task scheduler how many of them are launched in parallel.
If less than $q$ tasks are running simultaneously, tasks switch between cutters to advance the cutters with the currently smallest flow values.
If all $q$ tasks are running, tasks do not switch cutters.
This switching mechanism is light-weight due to its use of atomics, and incurs almost no overhead when $q$ tasks are available.

For every cutter, we store two atomic flags: an \emph{active} flag which indicates whether the cutter is not finished, and an \emph{acquired} flag which indicates whether a task currently holds this cutter.
In the beginning every cutter is active and not acquired.
In a running task, we acquire an active task with minimum flow, which has not been acquired.
If this is not possible, we terminate the task.
Otherwise, we check whether the cutter cannot improve expansion and deactivate it, if so.
If not, we push one unit of flow or derive a cut.
If we find a cut, we check whether it improves expansion and has at least $20\%$ of the nodes on the smaller side.
If so, we store the cut as the current best.
Finally we release the cutter and repeat.
Since we create $q$ tasks and we release the previous cutter before trying to acquire a new one, it is sufficient to try acquiring every active cutter once, and terminating the task if unsuccessful.

This scheme guarantees $\mathcal{O}(\frac{qcm}{k})$ span and $\mathcal{O}(qcm)$ work, for $k \leq q$ cores executing in parallel.
Note that due to the parallelization, cuts are not necessarily enumerated in order of increasing cut size and also dominated cuts may be reported.

%%%%%%%%%%%%%%%%%%%%%%%%%%%%%%%%%%%%%%%%%%
\section{Results}\label{sec:results}

In this section, we discuss our experimental setup and results. 

\subsection{Experimental Setup}
In Section~\ref{sec:parameterstudy} we perform a parameter study based on CCH performance, to obtain reasonable parameters for InertialFlowCutter.
The parameters are tuned for CCH performance, not top-level cuts.
Our remaining experiments follow the setup in~\cite{hs-gbpo-18}, comparing FlowCutter, KaHiP, Metis and Inertial Flow to InertialFlowCutter, regarding CCH performance as well as top-level cut sizes for different imbalances.
Our benchmark set consists of the road networks of Colorado, California and Nevada, the USA and Western Europe, see Table~\ref{table:benchmark_roadgraphs}, made available during the DIMACS implementation challenge on shortest paths~\cite{dgj-spndi-09}.

\begin{table}
	\begin{center}
	\begin{tabular}{l r r}
		\toprule
		Graph & $n$ & $m$ \\ \midrule
		Colorado & $436 \cdot 10^3$  & $10^6$ \\ \midrule
		California and Nevada & $1.9 \cdot 10^6$ & $4.6 \cdot 10^6$  \\ \midrule 
		USA & $24 \cdot 10^6$ & $57 \cdot 10^6$  \\ \midrule
		Europe & $18 \cdot 10^6$  & $44 \cdot 10^6$ \\
		\bottomrule
	\end{tabular}
	\caption{Benchmark road networks.}\label{table:benchmark_roadgraphs}
	\end{center}
\end{table}

The CCH performance experiments compare the different partitioners based on the time to compute a contraction order, the median running time of nine customization runs, the average time of $10^6$ random $s$-$t$ queries, as well as the criteria introduced in Section~\ref{sec:cch}.
Unless explicitly stated as parallel, all reported running times are sequential on an Intel Xeon E5-1630 v3 Haswell processor clocked at 3.7GHz with 10MB L3 cache and 128GB DDR4 RAM (2133 MHz).
We additionally report running times for computing contraction orders in parallel on a shared-memory machine with two 8-core Intel Xeon Gold 6144 Skylake CPUs, clocked at 3.5GHz with 24.75MB L3 cache and 192GB DDR4 RAM (2666 MHz).
InertialFlowCutter is implemented in C++ and the code is compiled with g++ version 8.2 with optimization level 3.
We use Intel's Threading Building Blocks library for shared-memory parallelism.
Our InertialFlowCutter implementation and evaluation scripts are available on GitHub~\cite{inertialflowcutter-github}.

\subsection{CCH Implementation}
We use the CCH implementation in RoutingKit~\cite{routingkit-github}.
There are different CCH customization and query variants.
We use basic customization with upper triangles instead of lower triangles, no witness searches, no precomputed triangles, no SSE, and no parallelization.
For queries we use elimination tree search.
There has been a recent, very simple improvement~\cite{bsw-rttau-18}, which drastically accelerates elimination tree search for short-range queries.
It is not implemented in RoutingKit but random $s$-$t$ queries tend to be long range, so the effect would be negligible for our experiments.

\subsection{Partitioner Implementations and Nested Dissection Setup}
In~\cite{hs-gbpo-18} the KaHiP versions 0.61 and 1.00 are used.
We did not re-run the preprocessing for those old versions of KaHiP but use the orders and running times of~\cite{hs-gbpo-18}.
The running times are comparable as the experiments are run on the same machine.
We add the latest KaHiP version 2.11, which is available on GitHub~\cite{kahip-github}.
For all three versions the  \emph{strong} preset of KaHiP is used.
We refer to the three KaHiP variants as K0.61, K1.00 and K2.11.
For the CCH order experiments we keep versions K0.61 and K1.00 but omit them for the top-level cut experiments because K2.11 is better for top-level cuts.

We use Metis 5.1.0, available from the authors' website~\cite{metis-website}, which we denote by M in our tables.

We use InertialFlowCutter with $\langle 4,8,12,16 \rangle$ directions and denote the configurations by IFC4, IFC8, IFC12, IFC16, respectively.

We use our own Inertial Flow implementation with the four directions proposed in~\cite{ss-obsrn-15}.
It is available at our repository~\cite{inertialflowcutter-github}.
Instead of Dinic algorithm~\cite{d-aspmf-70} we use Ford-Fulkerson, as preliminary experiments indicate it is faster.
Further, we filter source nodes that are only connected to other sources, and target nodes that are only connected to other targets.
Instead of sorting nodes along a direction, we partition the node-array such that the first and last $\alpha \cdot n$ nodes are the desired terminals, using \texttt{std::nth\_element}.
These optimizations reduce the running time from 1017 seconds~\cite{hs-gbpo-18} down to 450 seconds for a CCH order on Europe.
Additionally, we use flow-based interleaving on Inertial Flow.
This was already included in the Inertial Flow implementation used in~\cite{hs-gbpo-18}.
We denote Inertial Flow by I in our tables.

The original FlowCutter implementation used in~\cite{hs-gbpo-18} is available on GitHub\cite{flowcutter-github}.
We use a slightly modified version that has been adjusted to use Intel's Threading Building Blocks instead of OpenMP for optional parallelism.
All parallelism is disabled for FlowCutter in our experiments.
We use FlowCutter with $\langle 3,20,100 \rangle$ random source-target pairs and denote the variants by F3, F20, F100, respectively.

Implementations of Buffoon~\cite{ss-degp-12} and PUNCH~\cite{dgrw-gpnc-11} are not publicly available.
Therefore, these are not included in our experiments.
%According to~\cite{hs-gbpo-18} the quality of PUNCH is similar to KaHiP, based on similar top-level cuts on the Europe and USA road networks.

We now discuss the different node ordering setups used in the experiments.
Metis offers its own node ordering tool \texttt{ndmetis}, which we use.
For Inertial Flow, K1.00 and K2.11 we use a nested dissection implementation, which computes one edge cut per level and recurses until components are trees or cliques, which are solved directly.
Separators are derived by picking the nodes incident to one side of the edge cut.
For comparability with~\cite{hs-gbpo-18} and~\cite{dsw-cch-15} we use an older nested dissection implementation for K0.61, which, on every level repeatedly computes edge cuts until no smaller cut was found for ten consecutive iterations.
For InertialFlowCutter and FlowCutter, we employ the setup that was proposed for FlowCutter in~\cite{hs-gbpo-18} that has also been described in Section~\ref{sec:nested_dissection}.
Our nested dissection implementation is based on the implementation in the FlowCutter repository~\cite{flowcutter-github}.
We made minor changes and parallelized it, as described in Section~\ref{sec:parallelization}.

We tried to employ the special preprocessing techniques for KaHiP 2.11.
While this made order computation faster, the order quality was much worse regarding all criteria.

Starting with version 1.00, KaHiP includes a more sophisticated multilevel node separator algorithm~\cite{ss-amnsa-16}.
It was omitted from the experiments in~\cite{hs-gbpo-18} because it took 19 hours to compute an order for the small California graph, using one separator per level, and did not finish in reasonable time on the larger instances.
Therefore we still exclude it.

\subsection{Order Experiments}

\begin{table}[tp]
	\begin{tabular}{ *{2}{c} *{4}{r} *{3}{r} *{3}{r} }
            \toprule
            & & \multicolumn{4}{c}{Search Space} & CCH & & Up. & \multicolumn{3}{c}{Running times} \\
            \cmidrule(lr){3-6} \cmidrule(lr){10-12}
            & & \multicolumn{2}{c}{Nodes} & \multicolumn{2}{c}{Arcs {[}$\cdot10^{3}${]}} & Arcs &  \#Tri. & Tw. & Order & Cust. & Query \\
            \cmidrule(lr){3-4} \cmidrule(lr){5-6} 
            & & Avg. & Max.& Avg. & Max. & {[}$\cdot10^{6}${]} & {[}$\cdot10^{6}${]} & Bd. & {[}s{]} & {[}ms{]} & {[}$\mu$s{]}
        
\\ \midrule
\multirow{12}{*}{\begin{sideways}Col \end{sideways}} & M & 155.6 & 354 & 6.1 & 22.0 & 13.7 & 63.9 & 102 & \bfseries{1.8} & 58.8 & 21.1
\\
 & K0.61 & 135.1 & 357 & 4.6 & 21.6 & 16.7 & 72.4 & 103 & 3\,837.1 & 66.4 & 16.9
\\
 & K1.00 & 136.4 & 357 & 4.8 & 22.1 & 15.0 & 69.1 & 99 & 1\,052.4 & 62.0 & 17.1
\\
 & K2.11 & 135.1 & 363 & 4.7 & 22.8 & 14.9 & 68.4 & 100 & 924.6 & 61.8 & 16.9
\\
 & I & 151.3 & 543 & 6.2 & 37.7 & 15.0 & 73.9 & 119 & 3.0 & 63.7 & 20.1
\\
 & F3 & 127.2 & 277 & 4.1 & 14.4 & 12.8 & 47.4 & \bfseries{85} & 9.4 & 46.7 & 15.8
\\
 & F20 & 122.5 & 263 & \bfseries{3.8} & 13.8 & \bfseries{12.5} & 43.8 & 87 & 55.9 & 44.3 & 14.7
\\
 & F100 & \bfseries{122.3} & 263 & \bfseries{3.8} & 13.8 & \bfseries{12.5} & 43.7 & 87 & 274.5 & 44.4 & \bfseries{14.6}
\\
 & IFC4 & 123.2 & \bfseries{261} & 3.9 & \bfseries{13.7} & \bfseries{12.5} & 44.1 & 100 & 6.9 & 43.5 & 14.9
\\
 & IFC8 & 123.3 & \bfseries{261} & 3.9 & \bfseries{13.7} & \bfseries{12.5} & 43.9 & 100 & 12.9 & 43.4 & 14.9
\\
 & IFC12 & 123.1 & 263 & 3.9 & 14.0 & \bfseries{12.5} & 43.6 & 87 & 18.7 & 43.3 & 14.8
\\
 & IFC16 & 123.1 & 262 & 3.9 & 14.0 & \bfseries{12.5} & \bfseries{43.5} & 87 & 24.3 & \bfseries{43.2} & 14.8
\\
\midrule
\multirow{12}{*}{\begin{sideways}Cal \end{sideways}} & M & 275.5 & 543 & 17.3 & 53.2 & 65.0 & 364.1 & 180 & \bfseries{9.8} & 310.1 & 47.9
\\
 & K0.61 & 187.7 & 483 & 7.0 & 37.0 & 74.8 & 342.4 & 160 & 18\,659.3 & 316.4 & 24.9
\\
 & K1.00 & 184.9 & 471 & 6.8 & 37.9 & 69.5 & 334.4 & 143 & 6\,023.6 & 302.3 & 24.7
\\
 & K2.11 & 184.8 & 449 & 6.8 & 36.5 & 69.5 & 332.4 & 162 & 4\,374.9 & 300.8 & 24.7
\\
 & I & 191.4 & 605 & 7.1 & 53.4 & 68.8 & 341.3 & 161 & 16.0 & 301.7 & 25.4
\\
 & F3 & 178.8 & \bfseries{361} & 6.2 & \bfseries{24.9} & 59.2 & 235.4 & \bfseries{132} & 57.9 & 240.2 & 23.3
\\
 & F20 & 169.6 & 383 & \bfseries{5.6} & 26.3 & 58.0 & 218.5 & \bfseries{132} & 358.5 & 229.7 & 21.9
\\
 & F100 & 169.6 & 386 & \bfseries{5.6} & 26.3 & 58.0 & 218.3 & \bfseries{132} & 1\,759.2 & 229.9 & 21.9
\\
 & IFC4 & 170.0 & 380 & \bfseries{5.6} & 26.2 & 58.0 & 217.6 & \bfseries{132} & 42.3 & 225.1 & 21.7
\\
 & IFC8 & 169.8 & 380 & \bfseries{5.6} & 26.2 & 58.0 & 217.7 & \bfseries{132} & 79.0 & 225.1 & 21.7
\\
 & IFC12 & \bfseries{169.4} & 380 & \bfseries{5.6} & 26.2 & \bfseries{57.9} & \bfseries{217.2} & \bfseries{132} & 115.2 & \bfseries{224.8} & \bfseries{21.6}
\\
 & IFC16 & 170.2 & 381 & 5.7 & 26.2 & 58.0 & 218.4 & \bfseries{132} & 151.9 & 225.8 & 21.9
\\
\midrule
\multirow{12}{*}{\begin{sideways}Eur \end{sideways}} & M & 1\,167.3 & 1\,914 & 373.1 & 765.9 & 697.4 & 13\,238.1 & 828 & \bfseries{124.6} & 8\,302.3 & 645.3
\\
 & K0.61 & 638.6 & 1\,224 & 114.3 & 284.1 & 739.2 & 5\,782.5 & 482 & 213\,091.1 & 4\,464.5 & 229.1
\\
 & K1.00 & 652.5 & 1\,279 & 113.4 & 286.7 & 683.3 & 5\,745.4 & 451 & 242\,680.5 & 4\,169.7 & 223.7
\\
 & K2.11 & 652.6 & 1\,198 & 113.5 & 262.4 & 683.1 & 5\,637.7 & 449 & 49\,553.1 & 4\,125.0 & 224.1
\\
 & I & 732.6 & 1\,569 & 149.6 & 413.6 & 674.0 & 5\,897.3 & 516 & 450.3 & 4\,177.1 & 280.0
\\
 & F3 & 743.7 & 1\,156 & 138.1 & 283.7 & 602.1 & 5\,004.2 & 493 & 2\,227.9 & 3\,682.0 & 262.0
\\
 & F20 & 622.3 & 1\,142 & 106.6 & 262.1 & \bfseries{588.3} & 4\,624.1 & 454 & 16\,130.5 & 3\,527.4 & 211.4
\\
 & F100 & 615.5 & 1\,101 & 103.2 & \bfseries{237.2} & 588.4 & 4\,606.6 & 449 & 79\,176.6 & 3\,511.0 & 206.7
\\
 & IFC4 & 663.0 & \bfseries{1\,087} & 108.8 & 246.7 & 589.3 & 4\,644.9 & \bfseries{447} & 1\,245.7 & 3\,506.8 & 223.1
\\
 & IFC8 & \bfseries{608.6} & 1\,092 & \bfseries{102.1} & 246.7 & 588.6 & \bfseries{4\,587.1} & 454 & 2\,448.1 & 3\,508.5 & \bfseries{203.8}
\\
 & IFC12 & 611.1 & 1\,094 & 103.3 & 247.2 & 588.8 & 4\,627.5 & 454 & 3\,608.6 & 3\,511.9 & 205.7
\\
 & IFC16 & 609.5 & 1\,092 & 102.8 & 246.7 & 588.7 & 4\,616.6 & 454 & 4\,780.2 & \bfseries{3\,505.0} & 204.2
\\
\midrule
\multirow{12}{*}{\begin{sideways}USA \end{sideways}} & M & 1\,020.9 & 1\,763 & 273.6 & 666.7 & 861.7 & 12\,738.5 & 733 & \bfseries{171.9} & 7\,804.3 & 491.4
\\
 & K0.61 & 575.5 & 1\,041 & 71.3 & 185.0 & 979.0 & 7\,371.2 & 366 & 265\,567.3 & 5\,449.2 & 158.4
\\
 & K1.00 & 540.3 & 1\,063 & 62.3 & 208.1 & 887.4 & 6\,483.3 & 439 & 315\,942.6 & 4\,717.3 & 135.6
\\
 & K2.11 & 543.7 & 1\,015 & 63.2 & 190.2 & 887.4 & 6\,454.6 & 336 & 68\,828.1 & 4\,711.4 & 137.2
\\
 & I & 533.7 & 1\,371 & 62.0 & 290.9 & 887.9 & 6\,820.5 & 384 & 439.5 & 4\,821.1 & 135.5
\\
 & F3 & 512.0 & 929 & 57.5 & 163.0 & 758.9 & 4\,845.6 & 332 & 1\,813.0 & 3\,812.8 & 126.2
\\
 & F20 & 491.2 & 861 & 52.9 & 154.0 & 743.4 & 4\,425.2 & 312 & 11\,443.1 & 3\,610.6 & 119.0
\\
 & F100 & 491.1 & 864 & 52.8 & 153.9 & 743.6 & 4\,431.6 & 311 & 56\,934.7 & 3\,608.3 & 118.5
\\
 & IFC4 & 491.7 & 865 & 52.8 & \bfseries{153.4} & 743.1 & 4\,421.9 & \bfseries{310} & 1\,028.4 & 3\,608.2 & 118.4
\\
 & IFC8 & 491.4 & \bfseries{859} & 52.8 & \bfseries{153.4} & 743.0 & 4\,423.6 & 312 & 2\,022.9 & 3\,606.8 & 121.7
\\
 & IFC12 & \bfseries{490.7} & 865 & \bfseries{52.7} & \bfseries{153.4} & \bfseries{742.8} & \bfseries{4\,409.7} & 311 & 2\,977.3 & 3\,599.6 & \bfseries{118.1}
\\
 & IFC16 & 491.1 & 860 & 52.8 & \bfseries{153.4} & 742.9 & 4\,421.8 & 312 & 3\,938.5 & \bfseries{3\,592.8} & 118.2
\\
\bottomrule
\end{tabular}

	\caption{CCH order experiments.}
	\label{table:cch_orders}
\end{table}

In this section, we compare the different partitioners with respect to the quality of computed CCH orders and running time of the preprocessing.
Table~\ref{table:cch_orders} contains a large collection of metrics and measurements for the four road networks of California, Colorado, Europe and USA.

\paragraph{Quality}

Over all nodes $v$, we report the average and maximum number of ancestors in the elimination tree, as well as the number of arcs incident to the ancestors.
These metrics assess the search space sizes of an elimination tree query.
The query times in Table~\ref{table:cch_orders} are correlated with search space size, as expected.
The partitioner with the smallest average number of nodes and arcs in the search space always yields the fastest queries.
Further, we report the number of arcs in the CCH, \ie shortcut and original arcs, the number of triangles and an upper bound on the treewidth, which we obtain by using the CCH order as elimination ordering.
A CCH is essentially a chordal supergraph of the input.
Thus CCHs are closely related to tree decompositions and elimination orderings.
The relation between tree decompositions and Contraction Hierarchies is further explained in~\cite{sw-gfieo-15}.
A low treewidth usually corresponds to good performance with respect to the other metrics.
However, as the treewidth is defined by the largest bag in the tree decomposition which may depend on the size of few separators and disregards the size of all smaller separators, this is not always consistent.
In the context of shortest path queries, a better average is preferable to a slightly reduced maximum.

On the California and USA road networks, IFC12 yields the fastest queries and smallest average search space sizes, while on Europe IFC8 does.
On our smallest road network Colorado F100 is slightly ahead of the InertialFlowCutter variants by $0.2$ to $0.3$ microseconds query time.
IFC16 yields the fastest customization times on Colorado, Europe and the USA, while IFC12 yields the fastest customization times on California.
Customization times are correlated with the number of triangles.
However, on Europe and USA, the smallest number does not yield the fastest customizations.
Even though we take the median of nine runs, this may still be due to random fluctuations.

FlowCutter with at least 20 cutters, has slightly worse average search space sizes and query times than InertialFlowCutter on California and USA, but falls behind on Europe.
Thus InertialFlowCutter computes the best CCH orders, with FlowCutter close behind.
The different KaHiP variants and Inertial Flow compute the next best orders, while Metis is ranked last by a large margin.

The ratio between maximum and average search space size is most strongly pronounced for Inertial Flow.
This indicates that Inertial Flow works well for most separators but the quality degrades for a few.
InertialFlowCutter resolves this problem.

There is an interesting difference in the number of cutters necessary for good CCH orders with InertialFlowCutter and FlowCutter.
In~\cite{hs-gbpo-18}, F20 is the recommended configuration.
The performance differences between F20 and F100 are marginal (except on Europe).
However, using just 3 seems insufficient to get rid of bad random choices.

For the InertialFlowCutter variants, 4 cutters suffice most of the time.
The search space sizes, query times and customization times are very similar.
This is also confirmed by the top-level cut experiments in Section~\ref{sec:pareto_cuts}.
It seems the Inertial Flow guidance is sufficiently strong to eliminate bad random choices.
Again, only on Europe, the queries for IFC4 are slower, which is why we recommend using IFC8.
The better query running times justify the twice as long preprocessing.

Europe also stands out when comparing Inertial Flow query performances.
On Europe, Inertial Flow only beats Metis but on USA it beats all KaHiP versions and Metis.
The query performance difference of 57 microseconds between Inertial Flow and IFC4 on Europe suggests that the incremental cut computations of InertialFlowCutter make a significant difference and are worth the longer preprocessing times compared to Inertial Flow.

\paragraph{Preprocessing Time}

Previously, CCH performance came at the cost of high preprocessing time.
We compute better CCH orders than FlowCutter in a much shorter time.

KaHiP 0.61 and KaHiP 1.00 are by far the slowest.
KaHiP 2.11 is faster than F100, but slower than F20.
All InertialFlowCutter variants are faster than F20.
IFC8 and F3 have similar running times.
Metis is the fastest by a large margin and Inertial Flow is the second fastest.

The two old KaHiP versions are slow for different reasons.
As already mentioned K0.61 computes at least 10 cuts, as opposed to K1.00 and K2.11.
K1.00 is slow because the running time for top-level cuts with $\epsilon \geq 0.2$ increases unexpectedly, according to~\cite{hs-gbpo-18}.

\begin{table}
	\begin{center}
	\begin{tabular}{ll *{5}{r}}
\toprule
\multirow{2}{*}{Graph} &  & \multicolumn{5}{c}{Cores} \\
\cmidrule(lr){3-7}
 & & 1 & 2 & 4 & 8 & 16
\\
\midrule
\multirow{2}{*}{Col} & Time [s] & 11.6 & 6.1 & 3.3 & 2.1 & 1.7\\
 & Speedup & 1.0 & 1.9 & 3.5 & 5.5 & 6.8\\
\midrule
\multirow{2}{*}{Cal} & Time [s] & 71.5 & 36.7 & 19.2 & 11.3 & 7.2\\
 & Speedup & 1.0 & 1.9 & 3.7 & 6.3 & 9.9\\
\midrule
\multirow{2}{*}{Eur} & Time [s] & 2257.8 & 1160.0 & 600.7 & 334.2 & 241.8\\
 & Speedup & 1.0 & 1.9 & 3.8 & 6.8 & 9.3\\
\midrule
\multirow{2}{*}{USA} & Time [s] & 1869.5 & 947.9 & 497.0 & 275.5 & 173.2\\
 & Speedup & 1.0 & 2.0 & 3.8 & 6.8 & 10.8\\
\bottomrule
\end{tabular}

	\caption{Running times in seconds of IFC8, using up to 16 cores of the Skylake CPU.}
	\label{table:scalability}
	\end{center}
\end{table}

Using 16 cores and IFC8, we compute a CCH order of Europe in just 242 seconds, with 2258 seconds sequential running time on the Skylake machine.
This corresponds to a speedup of 9.3 over the sequential version.
See Table~\ref{table:scalability}.
Note that due to using 8 cutters, at most 8 threads work on a single separator.
Therefore, in particular for the top-level separator at most 8 of the 16 cores are used.
The top level separator alone needs about 50 seconds using 8 cores.
Due to unfortunate scheduling and unbalanced separators, it happens also at later stages that a single separator needs to be computed before any further tasks can be created.
Using 8 cores, we get a much better speedup of 6.8 for Europe, up to four cores we see an almost perfect speedup for all but the smallest road network.
This is because some cutters need less running time than others.
Thus there is actually less potential for parallelism than the number of cutters suggests.

\subsection{Pareto Cut Experiments}\label{sec:pareto_cuts}

\begin{table}[tp]
	
\newcolumntype{R}[1]{>{\raggedleft\arraybackslash}p{#1}}
\setlength\arraycolsep{50pt}
\setlength\tabcolsep{2pt}

\setlength\tabcolsep{3pt}
\setlength\mycolwidth{0.74cm}
\setlength\mysmallcolwidth{0.8cm}

\begin{tabular}{rR{\mycolwidth}R{\mycolwidth}R{\mycolwidth}R{\mycolwidth}R{\mycolwidth}R{\mycolwidth}R{\mycolwidth}R{\mycolwidth}R{\mycolwidth}R{\mycolwidth}R{\mycolwidth}R{\mycolwidth}R{\mycolwidth}R{\mycolwidth}R{\mycolwidth}R{\mycolwidth}}
\toprule
\multirow{2}{*}{\rotatebox[origin=c]{90}{$\max\epsilon$}}  & \multicolumn{8}{c}{Achieved $\epsilon$ [\%]} & \multicolumn{8}{c}{Cut Size}\\
\cmidrule(lr){2-9} \cmidrule(lr){10-17}
& IFC4& IFC8& IFC12& F3& F20& K2.11& M& I& IFC4& IFC8& IFC12& F3& F20& K2.11& M& I\\
\midrule
0& 0.0& 0.0& 0.0& 0.0& 0.0& \textcolor{red}{\cancel{1.0}}& --& 0.0& 60& 48& 48& 44& 44& \textcolor{red}{\cancel{35}}& --& 259\\
1& 0.4& 0.6& 0.2& 0.8& 0.8& 1.0& 0.0& 0.1& 47& 41& 38& 28& 28& 34& 37& 96\\
3& 2.8& 2.8& 1.9& 0.8& 0.8& 2.8& 0.0& 0.7& 36& 36& 36& 28& 28& 33& 57& 70\\
5& 4.3& 4.3& 4.3& 0.8& 0.8& 4.4& 2.7& 0.9& 28& 28& 28& 28& 28& 32& 39& 60\\
10& 8.9& 8.9& 8.9& 0.8& 9.1& 8.9& ${<0.1}$& 1.4& 22& 22& 22& 28& 22& 22& 43& 46\\
20& 11.6& 11.6& 11.6& 18.8& 18.8& 18.8& 16.7& 14.0& 20& 20& 20& 19& 19& 19& 230& 27\\
30& 27.6& 27.6& 27.6& 27.6& 27.6& 10.3& ${<0.1}$& 23.1& 14& 14& 14& 14& 14& 21& 44& 21\\
50& 40.6& 40.6& 40.6& 40.6& 40.6& 34.1& 44.3& 36.4& 12& 12& 12& 12& 12& 13& 22& 14\\
70& 57.6& 57.6& 57.6& 40.6& 57.6& 40.6& 41.2& 48.8& 11& 11& 11& 12& 11& 12& 1287& 12\\
90& 81.2& 89.0& 81.2& 83.5& 87.3& 89.4& 47.4& 81.5& 9& 6& 9& 11& 8& 5& 971& 9\\

\midrule
\multirow{2}{*}{\rotatebox[origin=c]{90}{$\max\epsilon$}}  & \multicolumn{8}{c}{Are sides connected?} & \multicolumn{8}{c}{Running Time [s]}\\
\cmidrule(lr){2-9} \cmidrule(lr){10-17}
& IFC4& IFC8& IFC12& F3& F20& K2.11& M& I& IFC4& IFC8& IFC12& F3& F20& K2.11& M& I\\
\midrule
0& $\bullet$& $\bullet$& $\bullet$& $\bullet$& $\bullet$& $\bullet$& --& $\circ$& 0.6& 0.8& 1.2& 1.6& 10.8& 2.1& --& 0.1\\
1& $\bullet$& $\bullet$& $\bullet$& $\bullet$& $\bullet$& $\bullet$& $\bullet$& $\circ$& 0.5& 0.7& 1.0& 1.3& 8.5& 2.4& 0.1& 0.1\\
3& $\bullet$& $\bullet$& $\bullet$& $\bullet$& $\bullet$& $\bullet$& $\bullet$& $\circ$& 0.4& 0.6& 0.9& 1.3& 8.5& 3.4& 0.1& 0.1\\
5& $\bullet$& $\bullet$& $\bullet$& $\bullet$& $\bullet$& $\bullet$& $\bullet$& $\circ$& 0.3& 0.5& 0.7& 1.3& 8.5& 4.6& 0.1& 0.1\\
10& $\bullet$& $\bullet$& $\bullet$& $\bullet$& $\bullet$& $\bullet$& $\bullet$& $\circ$& 0.2& 0.4& 0.6& 1.3& 7.1& 13.4& 0.1& 0.1\\
20& $\bullet$& $\bullet$& $\bullet$& $\bullet$& $\bullet$& $\bullet$& $\circ$& $\bullet$& 0.2& 0.4& 0.5& 0.9& 6.4& 26.2& 0.1& 0.1\\
30& $\bullet$& $\bullet$& $\bullet$& $\bullet$& $\bullet$& $\bullet$& $\bullet$& $\bullet$& 0.2& 0.3& 0.4& 0.7& 5.0& 26.7& 0.1& 0.1\\
50& $\bullet$& $\bullet$& $\bullet$& $\bullet$& $\bullet$& $\bullet$& $\bullet$& $\circ$& 0.1& 0.3& 0.3& 0.6& 4.4& 17.8& 0.1& 0.1\\
70& $\bullet$& $\bullet$& $\bullet$& $\bullet$& $\bullet$& $\bullet$& $\circ$& $\bullet$& 0.1& 0.2& 0.3& 0.6& 4.0& 43.7& 0.1& 0.1\\
90& $\bullet$& $\bullet$& $\bullet$& $\bullet$& $\bullet$& $\bullet$& $\circ$& $\bullet$& 0.1& 0.2& 0.3& 0.6& 3.0& 27.7& 0.1& 0.2\\

\bottomrule
\end{tabular}

	\caption{Colorado top-level cuts.}
	\label{table:toplevelcuts:col}
\end{table}

\begin{table}[tp]
	
\newcolumntype{R}[1]{>{\raggedleft\arraybackslash}p{#1}}
\setlength\arraycolsep{50pt}
\setlength\tabcolsep{2pt}

\setlength\tabcolsep{3pt}
\setlength\mycolwidth{0.74cm}
\setlength\mysmallcolwidth{0.8cm}

\begin{tabular}{rR{\mycolwidth}R{\mycolwidth}R{\mycolwidth}R{\mycolwidth}R{\mycolwidth}R{\mycolwidth}R{\mycolwidth}R{\mycolwidth}R{\mycolwidth}R{\mycolwidth}R{\mycolwidth}R{\mycolwidth}R{\mycolwidth}R{\mycolwidth}R{\mycolwidth}R{\mycolwidth}}
\toprule
\multirow{2}{*}{\rotatebox[origin=c]{90}{$\max\epsilon$}}  & \multicolumn{8}{c}{Achieved $\epsilon$ [\%]} & \multicolumn{8}{c}{Cut Size}\\
\cmidrule(lr){2-9} \cmidrule(lr){10-17}
& IFC4& IFC8& IFC12& F3& F20& K2.11& M& I& IFC4& IFC8& IFC12& F3& F20& K2.11& M& I\\
\midrule
0& 0.0& 0.0& 0.0& 0.0& 0.0& \textcolor{red}{\cancel{1.0}}& --& 0.0& 57& 46& 46& 80& 43& \textcolor{red}{\cancel{48}}& --& 306\\
1& 1.0& 1.0& 1.0& 0.2& 0.2& 0.2& 0.0& 0.6& 39& 35& 35& 61& 31& 31& 63& 93\\
3& 2.3& 2.3& 2.3& 2.4& 2.3& 2.3& ${<0.1}$& 1.1& 29& 29& 29& 50& 29& 29& 51& 64\\
5& 2.3& 2.3& 2.3& 4.3& 2.3& 2.3& ${<0.1}$& 1.6& 29& 29& 29& 34& 29& 29& 56& 62\\
10& 2.3& 2.3& 2.3& 5.3& 2.3& 2.3& 0.3& 0.6& 29& 29& 29& 29& 29& 29& 44& 37\\
20& 16.7& 16.7& 16.7& 5.3& 16.7& 16.7& ${<0.1}$& 2.7& 28& 28& 28& 29& 28& 28& 47& 29\\
30& 16.7& 16.7& 16.7& 5.3& 16.7& 2.3& ${<0.1}$& 5.5& 28& 28& 28& 29& 28& 29& 50& 29\\
50& 42.3& 42.3& 42.3& 5.3& 49.1& 2.3& 33.3& 40.8& 25& 25& 25& 29& 24& 29& 3118& 27\\
70& 42.3& 42.3& 42.3& 67.0& 49.1& 2.3& 41.2& 42.6& 25& 25& 25& 28& 24& 29& 3343& 26\\
90& 85.4& 85.4& 85.4& 90.0& 89.8& 49.1& 47.4& 85.6& 18& 18& 18& 15& 14& 24& 3040& 18\\

\midrule
\multirow{2}{*}{\rotatebox[origin=c]{90}{$\max\epsilon$}}  & \multicolumn{8}{c}{Are sides connected?} & \multicolumn{8}{c}{Running Time [s]}\\
\cmidrule(lr){2-9} \cmidrule(lr){10-17}
& IFC4& IFC8& IFC12& F3& F20& K2.11& M& I& IFC4& IFC8& IFC12& F3& F20& K2.11& M& I\\
\midrule
0& $\bullet$& $\bullet$& $\bullet$& $\bullet$& $\bullet$& $\bullet$& --& $\circ$& 2.5& 3.5& 5.0& 11.8& 58.9& 10.6& --& 0.3\\
1& $\bullet$& $\bullet$& $\bullet$& $\bullet$& $\bullet$& $\bullet$& $\bullet$& $\circ$& 1.9& 3.0& 4.3& 11.1& 49.6& 13.0& 0.7& 0.3\\
3& $\bullet$& $\bullet$& $\bullet$& $\bullet$& $\bullet$& $\bullet$& $\bullet$& $\circ$& 1.4& 2.5& 3.5& 10.1& 47.5& 22.7& 0.7& 0.3\\
5& $\bullet$& $\bullet$& $\bullet$& $\bullet$& $\bullet$& $\bullet$& $\bullet$& $\circ$& 1.4& 2.5& 3.5& 8.1& 47.5& 36.8& 0.7& 0.4\\
10& $\bullet$& $\bullet$& $\bullet$& $\bullet$& $\bullet$& $\bullet$& $\bullet$& $\circ$& 1.4& 2.5& 3.5& 7.2& 47.5& 74.5& 0.7& 0.4\\
20& $\bullet$& $\bullet$& $\bullet$& $\bullet$& $\bullet$& $\bullet$& $\bullet$& $\bullet$& 1.4& 2.3& 3.4& 7.2& 46.0& 104.0& 0.7& 0.5\\
30& $\bullet$& $\bullet$& $\bullet$& $\bullet$& $\bullet$& $\bullet$& $\bullet$& $\bullet$& 1.4& 2.3& 3.4& 7.2& 46.0& 172.6& 0.7& 0.6\\
50& $\bullet$& $\bullet$& $\bullet$& $\bullet$& $\bullet$& $\bullet$& $\circ$& $\circ$& 1.2& 2.0& 2.9& 7.2& 40.2& 210.8& 0.7& 1.1\\
70& $\bullet$& $\bullet$& $\bullet$& $\bullet$& $\bullet$& $\bullet$& $\circ$& $\circ$& 1.2& 2.0& 2.9& 6.9& 40.2& 227.6& 0.7& 1.5\\
90& $\bullet$& $\bullet$& $\bullet$& $\bullet$& $\bullet$& $\bullet$& $\circ$& $\bullet$& 0.8& 1.4& 2.1& 3.9& 23.5& 110.5& 0.7& 1.6\\

\bottomrule
\end{tabular}

	\caption{California and Nevada top-level cuts.}
	\label{table:toplevelcuts:cal}
\end{table}

\begin{table}[tp]
	
\newcolumntype{R}[1]{>{\raggedleft\arraybackslash}p{#1}}
\setlength\arraycolsep{50pt}
\setlength\tabcolsep{2pt}

\setlength\tabcolsep{3pt}
\setlength\mycolwidth{0.74cm}
\setlength\mysmallcolwidth{0.8cm}

\begin{tabular}{rR{\mycolwidth}R{\mycolwidth}R{\mycolwidth}R{\mycolwidth}R{\mycolwidth}R{\mycolwidth}R{\mycolwidth}R{\mycolwidth}R{\mycolwidth}R{\mycolwidth}R{\mycolwidth}R{\mycolwidth}R{\mycolwidth}R{\mycolwidth}R{\mycolwidth}R{\mycolwidth}}
\toprule
\multirow{2}{*}{\rotatebox[origin=c]{90}{$\max\epsilon$}}  & \multicolumn{8}{c}{Achieved $\epsilon$ [\%]} & \multicolumn{8}{c}{Cut Size}\\
\cmidrule(lr){2-9} \cmidrule(lr){10-17}
& IFC4& IFC8& IFC12& F3& F20& K2.11& M& I& IFC4& IFC8& IFC12& F3& F20& K2.11& M& I\\
\midrule
0& 0.0& 0.0& 0.0& 0.0& 0.0& \textcolor{red}{\cancel{1.0}}& --& 0.0& 311& 289& 288& 273& 271& \textcolor{red}{\cancel{148}}& --& 1578\\
1& 1.0& 1.0& 0.3& 0.7& 0.3& 1.0& ${<0.1}$& 0.3& 274& 274& 243& 246& 224& 148& 393& 417\\
3& 3.0& 3.0& 2.3& 0.7& 1.3& 2.6& ${<0.1}$& 0.4& 259& 241& 238& 246& 219& 130& 434& 340\\
5& 4.8& 4.8& 4.2& 4.6& 5.0& 2.9& ${<0.1}$& 0.2& 226& 226& 215& 211& 207& 129& 452& 299\\
10& 9.5& 9.5& 9.5& 9.5& 9.5& 7.9& ${<0.1}$& 0.2& 188& 188& 188& 188& 188& 112& 468& 284\\
20& 9.5& 9.5& 9.5& 9.5& 9.5& 7.8& ${<0.1}$& 7.5& 188& 188& 188& 188& 188& 113& 403& 229\\
30& 9.5& 9.5& 9.5& 9.5& 9.5& 26.8& ${<0.1}$& 9.1& 188& 188& 188& 188& 188& 104& 463& 202\\
50& 49.0& 49.0& 49.0& 9.5& 43.7& 8.2& 33.3& 9.5& 23& 23& 23& 188& 39& 111& 16151& 188\\
70& 49.0& 70.0& 49.0& 64.5& 67.5& 32.1& 41.2& 64.7& 23& 20& 23& 58& 22& 86& 23021& 38\\
90& 72.8& 72.8& 72.8& 72.8& 72.8& 72.8& 72.8& 72.8& 2& 2& 2& 2& 2& 2& 2& 2\\

\midrule
\multirow{2}{*}{\rotatebox[origin=c]{90}{$\max\epsilon$}}  & \multicolumn{8}{c}{Are sides connected?} & \multicolumn{8}{c}{Running Time [s]}\\
\cmidrule(lr){2-9} \cmidrule(lr){10-17}
& IFC4& IFC8& IFC12& F3& F20& K2.11& M& I& IFC4& IFC8& IFC12& F3& F20& K2.11& M& I\\
\midrule
0& $\bullet$& $\bullet$& $\bullet$& $\bullet$& $\bullet$& $\bullet$& --& $\circ$& 98.4& 183.3& 253.0& 503.6& 3240.0& 141.5& --& 5.1\\
1& $\bullet$& $\bullet$& $\bullet$& $\bullet$& $\bullet$& $\bullet$& $\circ$& $\circ$& 91.9& 179.3& 232.1& 475.1& 2965.1& 193.9& 8.1& 3.9\\
3& $\bullet$& $\bullet$& $\bullet$& $\bullet$& $\bullet$& $\bullet$& $\bullet$& $\circ$& 89.2& 169.0& 229.7& 475.1& 2930.9& 352.4& 8.1& 4.8\\
5& $\bullet$& $\bullet$& $\bullet$& $\bullet$& $\bullet$& $\bullet$& $\circ$& $\circ$& 81.9& 162.5& 216.3& 428.0& 2839.4& 639.1& 8.1& 6.4\\
10& $\bullet$& $\bullet$& $\bullet$& $\bullet$& $\bullet$& $\circ$& $\circ$& $\bullet$& 67.1& 138.5& 192.9& 390.1& 2647.1& 2256.7& 8.1& 11.2\\
20& $\bullet$& $\bullet$& $\bullet$& $\bullet$& $\bullet$& $\circ$& $\circ$& $\circ$& 67.1& 138.5& 192.9& 390.1& 2647.1& 3618.4& 8.1& 23.6\\
30& $\bullet$& $\bullet$& $\bullet$& $\bullet$& $\bullet$& $\circ$& $\circ$& $\bullet$& 67.1& 138.5& 192.9& 390.1& 2647.1& 2406.7& 8.1& 41.7\\
50& $\bullet$& $\bullet$& $\bullet$& $\bullet$& $\bullet$& $\circ$& $\circ$& $\bullet$& 10.7& 16.7& 24.1& 390.1& 613.5& 4233.7& 8.2& 86.5\\
70& $\bullet$& $\bullet$& $\bullet$& $\bullet$& $\bullet$& $\circ$& $\circ$& $\bullet$& 10.7& 14.9& 24.1& 124.1& 361.8& 3351.5& 8.2& 23.7\\
90& $\bullet$& $\bullet$& $\bullet$& $\bullet$& $\bullet$& $\bullet$& $\bullet$& $\bullet$& 4.3& 7.9& 11.9& 6.5& 49.1& 3353.0& 8.1& 4.8\\

\bottomrule
\end{tabular}

	\caption{Europe top-level cuts.}
	\label{table:toplevelcuts:europe}
\end{table}

\begin{table}[tp]
	
\newcolumntype{R}[1]{>{\raggedleft\arraybackslash}p{#1}}
\setlength\arraycolsep{50pt}
\setlength\tabcolsep{2pt}

\setlength\tabcolsep{3pt}
\setlength\mycolwidth{0.74cm}
\setlength\mysmallcolwidth{0.8cm}

\begin{tabular}{rR{\mycolwidth}R{\mycolwidth}R{\mycolwidth}R{\mycolwidth}R{\mycolwidth}R{\mycolwidth}R{\mycolwidth}R{\mycolwidth}R{\mycolwidth}R{\mycolwidth}R{\mycolwidth}R{\mycolwidth}R{\mycolwidth}R{\mycolwidth}R{\mycolwidth}R{\mycolwidth}}
\toprule
\multirow{2}{*}{\rotatebox[origin=c]{90}{$\max\epsilon$}}  & \multicolumn{8}{c}{Achieved $\epsilon$ [\%]} & \multicolumn{8}{c}{Cut Size}\\
\cmidrule(lr){2-9} \cmidrule(lr){10-17}
& IFC4& IFC8& IFC12& F3& F20& K2.11& M& I& IFC4& IFC8& IFC12& F3& F20& K2.11& M& I\\
\midrule
0& 0.0& 0.0& 0.0& 0.0& 0.0& \textcolor{red}{\cancel{0.8}}& --& 0.0& 115& 115& 115& 115& 115& \textcolor{red}{\cancel{118}}& --& 1579\\
1& 0.6& 0.6& 0.6& 0.6& 0.6& 0.5& 0.0& 0.4& 82& 82& 82& 82& 82& 94& 178& 406\\
3& 2.3& 2.3& 2.3& 2.3& 2.3& 2.4& ${<0.1}$& 0.1& 76& 76& 76& 76& 76& 73& 192& 257\\
5& 3.8& 3.8& 3.8& 3.8& 3.8& 3.8& 0.0& 0.1& 61& 61& 61& 61& 61& 61& 289& 186\\
10& 3.8& 3.8& 3.8& 3.8& 3.8& 3.8& ${<0.1}$& 3.2& 61& 61& 61& 61& 61& 61& 253& 81\\
20& 3.8& 3.8& 3.8& 3.8& 3.8& 3.8& ${<0.1}$& 3.9& 61& 61& 61& 61& 61& 61& 222& 61\\
30& 3.8& 3.8& 3.8& 3.8& 3.8& 3.8& ${<0.1}$& 3.9& 61& 61& 61& 61& 61& 61& 232& 61\\
50& 3.8& 3.8& 3.8& 3.8& 3.8& 3.8& 3.7& 3.9& 61& 61& 61& 61& 61& 61& 242& 61\\
70& 69.6& 69.6& 69.6& 69.6& 69.6& 3.8& 41.2& 66.5& 46& 46& 46& 46& 46& 61& 41976& 61\\
90& 69.6& 69.6& 69.6& 69.6& 69.6& 69.6& 47.4& 70.3& 46& 46& 46& 46& 46& 46& 45409& 46\\

\midrule
\multirow{2}{*}{\rotatebox[origin=c]{90}{$\max\epsilon$}}  & \multicolumn{8}{c}{Are sides connected?} & \multicolumn{8}{c}{Running Time [s]}\\
\cmidrule(lr){2-9} \cmidrule(lr){10-17}
& IFC4& IFC8& IFC12& F3& F20& K2.11& M& I& IFC4& IFC8& IFC12& F3& F20& K2.11& M& I\\
\midrule
0& $\bullet$& $\bullet$& $\bullet$& $\bullet$& $\bullet$& $\bullet$& --& $\circ$& 60.6& 102.0& 145.3& 246.9& 1963.9& 179.2& --& 6.5\\
1& $\bullet$& $\bullet$& $\bullet$& $\bullet$& $\bullet$& $\bullet$& $\bullet$& $\circ$& 43.7& 81.5& 117.4& 234.5& 1628.9& 246.5& 10.9& 5.1\\
3& $\bullet$& $\bullet$& $\bullet$& $\bullet$& $\bullet$& $\circ$& $\circ$& $\circ$& 40.4& 75.7& 108.2& 223.9& 1533.5& 691.4& 10.9& 5.5\\
5& $\bullet$& $\bullet$& $\bullet$& $\bullet$& $\bullet$& $\bullet$& $\bullet$& $\circ$& 31.8& 60.2& 86.0& 198.0& 1290.5& 1329.8& 10.9& 6.0\\
10& $\bullet$& $\bullet$& $\bullet$& $\bullet$& $\bullet$& $\bullet$& $\bullet$& $\bullet$& 31.8& 60.2& 86.0& 198.0& 1290.5& 1710.7& 10.9& 6.2\\
20& $\bullet$& $\bullet$& $\bullet$& $\bullet$& $\bullet$& $\bullet$& $\circ$& $\bullet$& 31.8& 60.2& 86.0& 198.0& 1290.5& 2983.5& 10.9& 9.3\\
30& $\bullet$& $\bullet$& $\bullet$& $\bullet$& $\bullet$& $\bullet$& $\bullet$& $\bullet$& 31.8& 60.2& 86.0& 198.0& 1290.5& 4891.8& 10.9& 16.6\\
50& $\bullet$& $\bullet$& $\bullet$& $\bullet$& $\bullet$& $\bullet$& $\circ$& $\bullet$& 31.8& 60.2& 86.0& 198.0& 1290.5& 5307.4& 10.9& 32.4\\
70& $\bullet$& $\bullet$& $\bullet$& $\bullet$& $\bullet$& $\bullet$& $\circ$& $\bullet$& 24.4& 44.7& 63.2& 154.2& 985.1& 5445.2& 11.2& 50.9\\
90& $\bullet$& $\bullet$& $\bullet$& $\bullet$& $\bullet$& $\bullet$& $\circ$& $\bullet$& 24.4& 44.7& 63.2& 154.2& 985.1& 10637.7& 11.2& 61.2\\

\bottomrule
\end{tabular}

	\caption{USA top-level cuts.}
	\label{table:toplevelcuts:usa}
\end{table}

For the top-level cut experiments, we permutate the nodes in preorder from a randomly selected start node, using the same start node for all partitioners.
Without randomization, reordering in preorder is the first step of every recursive call in the nested dissection implementation from~\cite{flowcutter-github}, in order to identify connected components and create subgraph copies with local node and arc identifiers.
Connected components are identified after deleting separators, and at the beginning, in case the graph is disconnected.
We chose to emulate this for the top-level cut experiments, to recreate the conditions of the nested dissection application.

In Tables~\ref{table:toplevelcuts:col}, \ref{table:toplevelcuts:cal}, \ref{table:toplevelcuts:europe} and \ref{table:toplevelcuts:usa} we report the found cuts for various values of $\epsilon$ for all road networks.
We use the partitioners KaHiP 2.11, IFC4, IFC8, IFC12, F3, F20, Metis, and Inertial Flow.
We also report the actually achieved imbalance, the running time and whether the sides of the cut are connected.
We report $\epsilon = 0.0$ only if perfect balance was achieved, otherwise if the rounded value is $0.0$, we report $< 0.1 \%$.
For none of the graphs, KaHIP was able to achieve perfect balance if perfect balance was desired.
We note this by crossing out the respective values.
This is due to our use of the KaHiP library interface that does not support enforcing balance.
Metis simply rejects $\epsilon = 0$, which is why we mark the corresponding entries with a dash.
Perfect balance is not actually useful for the application.
We solely include it to analyze the different Pareto cuts.

Note that for FlowCutter and InertialFlowCutter, the running time always includes the computation of all more imbalanced cuts, i.e., to generate the full set of cuts, only the running time of the perfectly balanced cut is needed while for all other partitioners, the sum of all reported running times is needed.

Concerning the performance, Metis wins but almost all reported cuts are larger than the cuts reported by FlowCutter, InertialFlowCutter and KaHiP.
Inertial Flow is also quite fast, but, due to its design, produces cuts that are much more balanced than desired and thus cannot achieve as small cuts as the other partitioners.

KaHIP achieves exceptionally small, highly balanced cuts on the Europe road network.
On the other road networks it is similar to or worse than F20 in terms of cut size.
This is due to the special geography of the Europe road network.
It excludes large parts of Eastern Europe, which is why there is a cut of size $2$ and $\epsilon = 72.8 \%$ imbalance that separates Norway, Sweden, and Finland from the rest of Europe.
For $\epsilon = 10\%$, KaHiP computes a cut with 112 edges, which separates the European mainland from the Iberian peninsula, Britain, Scandinavia minus Denmark, Italy and Austria~\cite{hs-gbpo-18}.
The alps separate Italy from the rest of Europe.
Britain is only connected via ferries, and the Iberian peninsula is separated by the Pyrénées.
One side of the cut is not connected because the only ferry between Britain and Scandinavia runs between Britain and Denmark.
FlowCutter is unable to find cuts with disconnected sides without a modified initialization.
By handpicking terminals for FlowCutter, a similar cut with only 87 edges and $15\%$ imbalance, which places Austria with the mainland instead, is found in~\cite{hs-gbpo-18}.
However, it turns out that the FlowCutter CCH order using the 87 edge cut as a top-level separator is not much better than plain FlowCutter.
This indicates that it does not matter at what level of recursion the different cuts are found.

For large imbalances, KaHIP seems unable to leverage the additional freedom to achieve the much smaller but more unbalanced cuts, like the ones reported by InertialFlowCutter and FlowCutter.
This has already been observed for previous versions of KaHIP~\cite{hs-gbpo-18}.
In terms of running time, KaHIP and F20 are the slowest algorithms.
InertialFlowCutter is in all three configurations an order of magnitude faster than F20.
Up to a maximum $\epsilon$ of $10\%$, the three variants report almost the same cuts.
Apart from the very imbalanced $\epsilon = 90\%$ cuts, the cuts are also at most one edge worse than F20.
Only for more balanced cuts, more cutters give a significant improvement.
Here, in particular on the Europe road network, F20 is also significantly better than InertialFlowCutter.
In the range between $\epsilon = 60\%$ and $\epsilon = 10\%$, which is most relevant for our application, there is thus no significant difference between F20 and InertialFlowCutter, regardless of the number of cutters.
This indicates that on the top level, the first four directions seem to cover most cuts already.
On the other hand, for highly balanced cuts, the geographic initialization does not help much, as can be seen from the much worse cuts for InertialFlowCutter.
Here, just having more cutters seems to help.

\subsection{Parameter Configuration}\label{sec:parameterstudy}
In this section, we tune the parameters $\alpha, \delta, \gamma_a, \gamma_o$ of InertialFlowCutter.
Our goal is to achieve much faster order computation without sacrificing CCH performance.
Recall that $\alpha$ is the fraction of nodes initially fixed on each side, $\delta$ is -- roughly speaking -- a stepsize, $\gamma_o$ is the threshold up to how many nodes on a side of the projection we perform bulk piercing, and similarly $\gamma_a$ for how many settled nodes on a side.
Table~\ref{table:parameter_study} shows a large variety of tested parameter combinations for InertialFlowCutter with 8 directions on the road network of Europe.
We select the parameter set $\alpha = 0.05, \delta = 0.05, \gamma_a = 0.4, \gamma_o = 0.25$ based on query performance.
The best entries per column are highlighted in bold.
Further, color shades are scaled between values in the columns.
Darker shades correspond to lower values, which are better for every measure.

First, we consider the top part of Table~\ref{table:parameter_study}, where we fix $\alpha$ to $0.05$ and try different combinations of $\delta, \gamma_o, \gamma_a$.
While the number of triangles and customization times are correlated, the top configurations for these measures are not the same; interestingly.
The variations in search space sizes, customization time (27ms) and query time (3$\mu$s) are marginal.
Therefore we settle on the configuration $\delta = 0.05, \gamma_o = 0.25, \gamma_a = 0.4$, which simultaneously yields the fastest query and order times.

In the bottom part of Table~\ref{table:parameter_study} we try different values of $\alpha$ with the best choices for the other parameters.
As expected, larger values for $\alpha$ accelerate order computation and slightly slow down queries.

In summary, InertialFlowCutter is relatively robust to parameter choices other than for $\alpha$, which means users do not need to invest much effort on parameter tuning.

\begin{table}[tp]
	\begin{tabular}{ *{4}{c} *{4}{r} *{3}{r} *{3}{r} }
            \toprule
            & & & & \multicolumn{4}{c}{Search Space} & CCH & & Up. & \multicolumn{3}{c}{Running times} \\
            \cmidrule(lr){5-8} \cmidrule(lr){12-14}
            \multicolumn{4}{c}{Configuration} & \multicolumn{2}{c}{Nodes} & \multicolumn{2}{c}{Arcs {[}$\cdot10^{3}${]}} & Arcs &  \#Tri. & Tw. & Order & Cust. & Query \\
            \cmidrule(lr){1-4} \cmidrule(lr){5-6} \cmidrule(lr){7-8}
            $\alpha$ & $\delta$ & $\gamma_a$ & $\gamma_o$ & Avg. & Max.& Avg. & Max. & {[}$\cdot10^{6}${]} & {[}$\cdot10^{6}${]} & Bd. & {[}s{]} & {[}ms{]} & {[}$\mu$s{]}\\
            \midrule
        
0.05 & 0.05 & 0.3 & 0.1 & \cellcolor{cyan!100.0}610.2 & \cellcolor{cyan!100.0}\bfseries{1\,092} & \cellcolor{cyan!100.0}102.8 & \cellcolor{cyan!100.0}248.9 & \cellcolor{cyan!90.0}588.7 & \cellcolor{cyan!90.0}4\,586.6 & \cellcolor{cyan!90.0}454 & \cellcolor{cyan!10.0}2\,933 & \cellcolor{cyan!90.0}3\,388 & \cellcolor{cyan!100.0}204.1\\
0.05 & 0.05 & 0.3 & 0.15 & \cellcolor{cyan!100.0}608.6 & \cellcolor{cyan!100.0}1\,093 & \cellcolor{cyan!100.0}102.2 & \cellcolor{cyan!100.0}248.9 & \cellcolor{cyan!100.0}\bfseries{588.6} & \cellcolor{cyan!100.0}4\,578.1 & \cellcolor{cyan!90.0}454 & \cellcolor{cyan!30.0}2\,644 & \cellcolor{cyan!100.0}\bfseries{3\,380} & \cellcolor{cyan!100.0}203.2\\
0.05 & 0.05 & 0.35 & 0.15 & \cellcolor{cyan!100.0}608.6 & \cellcolor{cyan!100.0}1\,093 & \cellcolor{cyan!100.0}102.2 & \cellcolor{cyan!100.0}248.9 & \cellcolor{cyan!100.0}\bfseries{588.6} & \cellcolor{cyan!100.0}\bfseries{4\,577.8} & \cellcolor{cyan!90.0}454 & \cellcolor{cyan!30.0}2\,655 & \cellcolor{cyan!90.0}3\,385 & \cellcolor{cyan!100.0}203.2\\
0.05 & 0.05 & 0.3 & 0.2 & \cellcolor{cyan!100.0}610.6 & \cellcolor{cyan!100.0}1\,096 & \cellcolor{cyan!90.0}103.0 & \cellcolor{cyan!100.0}248.9 & \cellcolor{cyan!60.0}588.9 & \cellcolor{cyan!50.0}4\,621.6 & \cellcolor{cyan!90.0}454 & \cellcolor{cyan!40.0}2\,505 & \cellcolor{cyan!50.0}3\,400 & \cellcolor{cyan!100.0}204.1\\
0.05 & 0.05 & 0.35 & 0.2 & \cellcolor{cyan!100.0}610.5 & \cellcolor{cyan!90.0}1\,098 & \cellcolor{cyan!90.0}103.0 & \cellcolor{cyan!100.0}\bfseries{246.4} & \cellcolor{cyan!60.0}588.9 & \cellcolor{cyan!50.0}4\,620.8 & \cellcolor{cyan!90.0}454 & \cellcolor{cyan!40.0}2\,487 & \cellcolor{cyan!60.0}3\,396 & \cellcolor{cyan!100.0}203.9\\
0.05 & 0.05 & 0.4 & 0.2 & \cellcolor{cyan!100.0}610.3 & \cellcolor{cyan!90.0}1\,098 & \cellcolor{cyan!100.0}102.9 & \cellcolor{cyan!100.0}246.7 & \cellcolor{cyan!60.0}588.9 & \cellcolor{cyan!40.0}4\,622.2 & \cellcolor{cyan!90.0}454 & \cellcolor{cyan!40.0}2\,495 & \cellcolor{cyan!50.0}3\,400 & \cellcolor{cyan!100.0}204.4\\
0.05 & 0.05 & 0.3 & 0.25 & \cellcolor{cyan!100.0}610.5 & \cellcolor{cyan!100.0}1\,096 & \cellcolor{cyan!90.0}103.0 & \cellcolor{cyan!100.0}248.9 & \cellcolor{cyan!60.0}588.9 & \cellcolor{cyan!30.0}4\,630.8 & \cellcolor{cyan!90.0}454 & \cellcolor{cyan!40.0}2\,476 & \cellcolor{cyan!40.0}3\,404 & \cellcolor{cyan!100.0}204.4\\
0.05 & 0.05 & 0.35 & 0.25 & \cellcolor{cyan!100.0}610.6 & \cellcolor{cyan!100.0}\bfseries{1\,092} & \cellcolor{cyan!90.0}103.1 & \cellcolor{cyan!100.0}\bfseries{246.4} & \cellcolor{cyan!60.0}588.9 & \cellcolor{cyan!30.0}4\,629.6 & \cellcolor{cyan!90.0}454 & \cellcolor{cyan!40.0}2\,464 & \cellcolor{cyan!40.0}3\,403 & \cellcolor{cyan!100.0}204.3\\
0.05 & 0.05 & 0.4 & 0.25 & \cellcolor{cyan!100.0}608.6 & \cellcolor{cyan!100.0}\bfseries{1\,092} & \cellcolor{cyan!100.0}102.1 & \cellcolor{cyan!100.0}246.7 & \cellcolor{cyan!100.0}\bfseries{588.6} & \cellcolor{cyan!90.0}4\,587.1 & \cellcolor{cyan!90.0}454 & \cellcolor{cyan!40.0}2\,448 & \cellcolor{cyan!50.0}3\,401 & \cellcolor{cyan!100.0}\bfseries{202.9}\\
0.05 & 0.05 & 0.35 & 0.3 & \cellcolor{cyan!100.0}610.6 & \cellcolor{cyan!100.0}\bfseries{1\,092} & \cellcolor{cyan!90.0}103.0 & \cellcolor{cyan!100.0}\bfseries{246.4} & \cellcolor{cyan!70.0}588.8 & \cellcolor{cyan!40.0}4\,628.0 & \cellcolor{cyan!90.0}454 & \cellcolor{cyan!40.0}2\,457 & \cellcolor{cyan!60.0}3\,396 & \cellcolor{cyan!100.0}204.2\\
0.05 & 0.05 & 0.4 & 0.3 & \cellcolor{cyan!100.0}609.5 & \cellcolor{cyan!100.0}\bfseries{1\,092} & \cellcolor{cyan!100.0}102.9 & \cellcolor{cyan!100.0}246.7 & \cellcolor{cyan!90.0}588.7 & \cellcolor{cyan!40.0}4\,625.6 & \cellcolor{cyan!90.0}454 & \cellcolor{cyan!40.0}2\,445 & \cellcolor{cyan!60.0}3\,398 & \cellcolor{cyan!100.0}203.7\\
0.05 & 0.05 & 0.4 & 0.35 & \cellcolor{cyan!100.0}609.6 & \cellcolor{cyan!100.0}1\,094 & \cellcolor{cyan!100.0}102.9 & \cellcolor{cyan!100.0}246.7 & \cellcolor{cyan!70.0}588.8 & \cellcolor{cyan!40.0}4\,626.7 & \cellcolor{cyan!90.0}454 & \cellcolor{cyan!40.0}2\,445 & \cellcolor{cyan!40.0}3\,404 & \cellcolor{cyan!100.0}203.9\\
0.05 & 0.1 & 0.3 & 0.1 & \cellcolor{cyan!100.0}610.3 & \cellcolor{cyan!100.0}\bfseries{1\,092} & \cellcolor{cyan!100.0}102.9 & \cellcolor{cyan!100.0}248.9 & \cellcolor{cyan!70.0}588.8 & \cellcolor{cyan!80.0}4\,594.8 & \cellcolor{cyan!90.0}454 & \cellcolor{cyan!10.0}2\,904 & \cellcolor{cyan!80.0}3\,391 & \cellcolor{cyan!100.0}204.4\\
0.05 & 0.1 & 0.3 & 0.15 & \cellcolor{cyan!100.0}610.7 & \cellcolor{cyan!50.0}1\,116 & \cellcolor{cyan!90.0}103.1 & \cellcolor{cyan!100.0}248.9 & \cellcolor{cyan!70.0}588.8 & \cellcolor{cyan!70.0}4\,603.4 & \cellcolor{cyan!90.0}454 & \cellcolor{cyan!30.0}2\,595 & \cellcolor{cyan!20.0}3\,413 & \cellcolor{cyan!100.0}204.5\\
0.05 & 0.1 & 0.35 & 0.15 & \cellcolor{cyan!100.0}608.5 & \cellcolor{cyan!50.0}1\,116 & \cellcolor{cyan!100.0}102.2 & \cellcolor{cyan!100.0}248.9 & \cellcolor{cyan!90.0}588.7 & \cellcolor{cyan!90.0}4\,586.1 & \cellcolor{cyan!90.0}454 & \cellcolor{cyan!30.0}2\,589 & \cellcolor{cyan!60.0}3\,399 & \cellcolor{cyan!70.0}208.2\\
0.05 & 0.1 & 0.3 & 0.2 & \cellcolor{cyan!90.0}612.4 & \cellcolor{cyan!100.0}1\,094 & \cellcolor{cyan!90.0}103.8 & \cellcolor{cyan!100.0}248.9 & \cellcolor{cyan!60.0}588.9 & \cellcolor{cyan!40.0}4\,628.5 & \cellcolor{cyan!90.0}454 & \cellcolor{cyan!40.0}2\,508 & \cellcolor{cyan!50.0}3\,402 & \cellcolor{cyan!90.0}205.8\\
0.05 & 0.1 & 0.35 & 0.2 & \cellcolor{cyan!100.0}610.1 & \cellcolor{cyan!100.0}1\,093 & \cellcolor{cyan!100.0}102.8 & \cellcolor{cyan!100.0}\bfseries{246.4} & \cellcolor{cyan!70.0}588.8 & \cellcolor{cyan!60.0}4\,611.8 & \cellcolor{cyan!90.0}454 & \cellcolor{cyan!40.0}2\,480 & \cellcolor{cyan!60.0}3\,396 & \cellcolor{cyan!100.0}203.8\\
0.05 & 0.1 & 0.4 & 0.2 & \cellcolor{cyan!100.0}610.3 & \cellcolor{cyan!100.0}1\,093 & \cellcolor{cyan!100.0}102.9 & \cellcolor{cyan!100.0}246.7 & \cellcolor{cyan!60.0}588.9 & \cellcolor{cyan!50.0}4\,616.6 & \cellcolor{cyan!90.0}454 & \cellcolor{cyan!40.0}2\,489 & \cellcolor{cyan!60.0}3\,398 & \cellcolor{cyan!100.0}203.9\\
0.05 & 0.1 & 0.3 & 0.25 & \cellcolor{cyan!90.0}612.3 & \cellcolor{cyan!100.0}1\,094 & \cellcolor{cyan!90.0}103.7 & \cellcolor{cyan!100.0}248.9 & \cellcolor{cyan!60.0}588.9 & \cellcolor{cyan!40.0}4\,628.7 & \cellcolor{cyan!90.0}454 & \cellcolor{cyan!40.0}2\,516 & \cellcolor{cyan!50.0}3\,400 & \cellcolor{cyan!90.0}205.6\\
0.05 & 0.1 & 0.35 & 0.25 & \cellcolor{cyan!100.0}610.1 & \cellcolor{cyan!90.0}1\,099 & \cellcolor{cyan!100.0}102.8 & \cellcolor{cyan!100.0}\bfseries{246.4} & \cellcolor{cyan!70.0}588.8 & \cellcolor{cyan!60.0}4\,610.3 & \cellcolor{cyan!90.0}454 & \cellcolor{cyan!40.0}2\,497 & \cellcolor{cyan!70.0}3\,392 & \cellcolor{cyan!100.0}204.1\\
0.05 & 0.1 & 0.4 & 0.25 & \cellcolor{cyan!100.0}608.2 & \cellcolor{cyan!90.0}1\,099 & \cellcolor{cyan!100.0}102.0 & \cellcolor{cyan!100.0}246.7 & \cellcolor{cyan!100.0}\bfseries{588.6} & \cellcolor{cyan!100.0}4\,579.1 & \cellcolor{cyan!90.0}454 & \cellcolor{cyan!40.0}2\,478 & \cellcolor{cyan!90.0}3\,388 & \cellcolor{cyan!100.0}203.4\\
0.05 & 0.1 & 0.35 & 0.3 & \cellcolor{cyan!100.0}610.2 & \cellcolor{cyan!100.0}1\,093 & \cellcolor{cyan!100.0}102.9 & \cellcolor{cyan!100.0}\bfseries{246.4} & \cellcolor{cyan!70.0}588.8 & \cellcolor{cyan!50.0}4\,614.8 & \cellcolor{cyan!90.0}454 & \cellcolor{cyan!40.0}2\,489 & \cellcolor{cyan!60.0}3\,396 & \cellcolor{cyan!100.0}204.0\\
0.05 & 0.1 & 0.4 & 0.3 & \cellcolor{cyan!100.0}609.5 & \cellcolor{cyan!100.0}1\,093 & \cellcolor{cyan!100.0}102.9 & \cellcolor{cyan!100.0}246.7 & \cellcolor{cyan!70.0}588.8 & \cellcolor{cyan!40.0}4\,622.7 & \cellcolor{cyan!90.0}454 & \cellcolor{cyan!40.0}2\,482 & \cellcolor{cyan!70.0}3\,395 & \cellcolor{cyan!100.0}203.8\\
0.05 & 0.1 & 0.4 & 0.35 & \cellcolor{cyan!100.0}609.6 & \cellcolor{cyan!100.0}1\,095 & \cellcolor{cyan!100.0}102.9 & \cellcolor{cyan!100.0}246.7 & \cellcolor{cyan!70.0}588.8 & \cellcolor{cyan!40.0}4\,623.1 & \cellcolor{cyan!90.0}454 & \cellcolor{cyan!40.0}2\,475 & \cellcolor{cyan!60.0}3\,397 & \cellcolor{cyan!100.0}203.6\\
0.05 & 0.15 & 0.3 & 0.1 & \cellcolor{cyan!100.0}610.3 & \cellcolor{cyan!100.0}\bfseries{1\,092} & \cellcolor{cyan!100.0}102.9 & \cellcolor{cyan!100.0}248.9 & \cellcolor{cyan!70.0}588.8 & \cellcolor{cyan!80.0}4\,594.8 & \cellcolor{cyan!90.0}454 & \cellcolor{cyan!10.0}2\,906 & \cellcolor{cyan!60.0}3\,396 & \cellcolor{cyan!90.0}204.8\\
0.05 & 0.15 & 0.3 & 0.15 & \cellcolor{cyan!100.0}610.7 & \cellcolor{cyan!50.0}1\,116 & \cellcolor{cyan!90.0}103.1 & \cellcolor{cyan!100.0}248.9 & \cellcolor{cyan!70.0}588.8 & \cellcolor{cyan!70.0}4\,603.0 & \cellcolor{cyan!90.0}454 & \cellcolor{cyan!30.0}2\,572 & \cellcolor{cyan!70.0}3\,393 & \cellcolor{cyan!100.0}204.5\\
0.05 & 0.15 & 0.35 & 0.15 & \cellcolor{cyan!100.0}608.6 & \cellcolor{cyan!50.0}1\,116 & \cellcolor{cyan!100.0}102.2 & \cellcolor{cyan!100.0}248.9 & \cellcolor{cyan!90.0}588.7 & \cellcolor{cyan!90.0}4\,588.0 & \cellcolor{cyan!90.0}454 & \cellcolor{cyan!30.0}2\,568 & \cellcolor{cyan!60.0}3\,396 & \cellcolor{cyan!100.0}203.1\\
0.05 & 0.15 & 0.3 & 0.2 & \cellcolor{cyan!90.0}612.6 & \cellcolor{cyan!50.0}1\,116 & \cellcolor{cyan!90.0}103.8 & \cellcolor{cyan!100.0}248.9 & \cellcolor{cyan!40.0}589.0 & \cellcolor{cyan!20.0}4\,637.0 & \cellcolor{cyan!90.0}454 & \cellcolor{cyan!40.0}2\,523 & \cellcolor{cyan!30.0}3\,407 & \cellcolor{cyan!90.0}205.9\\
0.05 & 0.15 & 0.35 & 0.2 & \cellcolor{cyan!100.0}610.3 & \cellcolor{cyan!60.0}1\,114 & \cellcolor{cyan!100.0}102.9 & \cellcolor{cyan!100.0}246.6 & \cellcolor{cyan!60.0}588.9 & \cellcolor{cyan!50.0}4\,617.3 & \cellcolor{cyan!90.0}454 & \cellcolor{cyan!40.0}2\,494 & \cellcolor{cyan!50.0}3\,400 & \cellcolor{cyan!100.0}204.4\\
0.05 & 0.15 & 0.4 & 0.2 & \cellcolor{cyan!100.0}610.5 & \cellcolor{cyan!60.0}1\,114 & \cellcolor{cyan!100.0}102.9 & \cellcolor{cyan!100.0}246.7 & \cellcolor{cyan!60.0}588.9 & \cellcolor{cyan!40.0}4\,622.6 & \cellcolor{cyan!90.0}454 & \cellcolor{cyan!40.0}2\,504 & \cellcolor{cyan!50.0}3\,402 & \cellcolor{cyan!100.0}204.5\\
0.05 & 0.15 & 0.3 & 0.25 & \cellcolor{cyan!90.0}612.6 & \cellcolor{cyan!90.0}1\,100 & \cellcolor{cyan!90.0}103.8 & \cellcolor{cyan!100.0}248.9 & \cellcolor{cyan!20.0}589.1 & \cellcolor{cyan!10.0}4\,644.5 & \cellcolor{cyan!90.0}454 & \cellcolor{cyan!40.0}2\,521 & \cellcolor{cyan!30.0}3\,406 & \cellcolor{cyan!90.0}205.7\\
0.05 & 0.15 & 0.35 & 0.25 & \cellcolor{cyan!100.0}610.3 & \cellcolor{cyan!90.0}1\,100 & \cellcolor{cyan!100.0}102.9 & \cellcolor{cyan!100.0}\bfseries{246.4} & \cellcolor{cyan!60.0}588.9 & \cellcolor{cyan!50.0}4\,619.2 & \cellcolor{cyan!90.0}454 & \cellcolor{cyan!40.0}2\,507 & \cellcolor{cyan!60.0}3\,398 & \cellcolor{cyan!100.0}203.9\\
0.05 & 0.15 & 0.4 & 0.25 & \cellcolor{cyan!100.0}608.8 & \cellcolor{cyan!90.0}1\,100 & \cellcolor{cyan!100.0}102.1 & \cellcolor{cyan!100.0}246.7 & \cellcolor{cyan!90.0}588.7 & \cellcolor{cyan!90.0}4\,588.6 & \cellcolor{cyan!90.0}454 & \cellcolor{cyan!40.0}2\,489 & \cellcolor{cyan!70.0}3\,393 & \cellcolor{cyan!100.0}203.1\\
0.05 & 0.15 & 0.35 & 0.3 & \cellcolor{cyan!100.0}610.4 & \cellcolor{cyan!90.0}1\,100 & \cellcolor{cyan!100.0}102.9 & \cellcolor{cyan!100.0}\bfseries{246.4} & \cellcolor{cyan!60.0}588.9 & \cellcolor{cyan!40.0}4\,624.2 & \cellcolor{cyan!90.0}454 & \cellcolor{cyan!40.0}2\,506 & \cellcolor{cyan!50.0}3\,401 & \cellcolor{cyan!100.0}204.2\\
0.05 & 0.15 & 0.4 & 0.3 & \cellcolor{cyan!100.0}609.6 & \cellcolor{cyan!90.0}1\,100 & \cellcolor{cyan!100.0}102.9 & \cellcolor{cyan!100.0}246.7 & \cellcolor{cyan!70.0}588.8 & \cellcolor{cyan!40.0}4\,626.9 & \cellcolor{cyan!90.0}454 & \cellcolor{cyan!40.0}2\,495 & \cellcolor{cyan!70.0}3\,393 & \cellcolor{cyan!100.0}203.8\\
0.05 & 0.15 & 0.4 & 0.35 & \cellcolor{cyan!100.0}609.7 & \cellcolor{cyan!90.0}1\,100 & \cellcolor{cyan!100.0}102.9 & \cellcolor{cyan!100.0}246.7 & \cellcolor{cyan!60.0}588.9 & \cellcolor{cyan!30.0}4\,634.3 & \cellcolor{cyan!90.0}454 & \cellcolor{cyan!40.0}2\,492 & \cellcolor{cyan!30.0}3\,408 & \cellcolor{cyan!100.0}203.9\\
  \midrule
0.01 & 0.05 & 0.4 & 0.25 & \cellcolor{cyan!100.0}609.0 & \cellcolor{cyan!100.0}1\,095 & \cellcolor{cyan!100.0}102.4 & \cellcolor{cyan!100.0}248.5 & \cellcolor{cyan!90.0}588.7 & \cellcolor{cyan!80.0}4\,596.3 & \cellcolor{cyan!90.0}454 & \cellcolor{cyan!10.0}2\,817 & \cellcolor{cyan!90.0}3\,387 & \cellcolor{cyan!100.0}203.6\\
0.025 & 0.05 & 0.4 & 0.25 & \cellcolor{cyan!100.0}\bfseries{607.6} & \cellcolor{cyan!100.0}1\,095 & \cellcolor{cyan!100.0}\bfseries{101.9} & \cellcolor{cyan!100.0}248.5 & \cellcolor{cyan!100.0}\bfseries{588.6} & \cellcolor{cyan!90.0}4\,585.2 & \cellcolor{cyan!90.0}454 & \cellcolor{cyan!30.0}2\,658 & \cellcolor{cyan!70.0}3\,394 & \cellcolor{cyan!100.0}203.1\\
0.075 & 0.05 & 0.4 & 0.25 & \cellcolor{cyan!50.0}633.0 & \cellcolor{cyan!20.0}1\,131 & \cellcolor{cyan!20.0}110.8 & \cellcolor{cyan!40.0}265.8 & \cellcolor{cyan!60.0}588.9 & \cellcolor{cyan!20.0}4\,638.9 & \cellcolor{cyan!100.0}\bfseries{450} & \cellcolor{cyan!50.0}2\,402 & \cellcolor{cyan!10.0}3\,416 & \cellcolor{cyan!20.0}216.1\\
0.1 & 0.05 & 0.4 & 0.25 & \cellcolor{cyan!30.0}641.9 & \cellcolor{cyan!0.0}1\,140 & \cellcolor{cyan!0.0}112.7 & \cellcolor{cyan!0.0}274.9 & \cellcolor{cyan!40.0}589.0 & \cellcolor{cyan!0.0}4\,650.9 & \cellcolor{cyan!100.0}451 & \cellcolor{cyan!60.0}2\,182 & \cellcolor{cyan!10.0}3\,417 & \cellcolor{cyan!0.0}219.0\\
0.125 & 0.05 & 0.4 & 0.25 & \cellcolor{cyan!10.0}651.5 & \cellcolor{cyan!50.0}1\,118 & \cellcolor{cyan!70.0}106.2 & \cellcolor{cyan!40.0}263.8 & \cellcolor{cyan!20.0}589.1 & \cellcolor{cyan!50.0}4\,618.0 & \cellcolor{cyan!0.0}475 & \cellcolor{cyan!90.0}1\,800 & \cellcolor{cyan!90.0}3\,386 & \cellcolor{cyan!50.0}211.9\\
0.15 & 0.05 & 0.4 & 0.25 & \cellcolor{cyan!10.0}651.6 & \cellcolor{cyan!70.0}1\,108 & \cellcolor{cyan!70.0}106.2 & \cellcolor{cyan!50.0}263.3 & \cellcolor{cyan!0.0}589.2 & \cellcolor{cyan!50.0}4\,616.1 & \cellcolor{cyan!0.0}475 & \cellcolor{cyan!100.0}\bfseries{1\,656} & \cellcolor{cyan!80.0}3\,390 & \cellcolor{cyan!20.0}216.6\\
\bottomrule
\end{tabular}

	\caption{CCH performance of different parameter configurations of IFC8 on Europe. Bold values are the best in their category. Darkness of shading indicates better values.}
	\label{table:parameter_study}
\end{table}

%%%%%%%%%%%%%%%%%%%%%%%%%%%%%%%%%%%%%%%%%%
\section{Discussion}\label{sec:discussion}
%Authors should discuss the results and how they can be interpreted in perspective of previous studies and of the working hypotheses. The findings and their implications should be discussed in the broadest context possible. Future research directions may also be highlighted.

We have presented InertialFlowCutter, an algorithm that exploits geographical information to quickly compute high-quality bipartitions of road networks.
Our experiments show that we are able to compute nested dissection orders as used for CCHs more than six times faster than the previous state-of-the-art algorithm, FlowCutter.
Using 16 cores, we can compute a nested dissection order of the Europe road network in four minutes.
This makes CCHs even more attractive to be applied in practice.

An open question is how to transfer the ideas of large initial terminal node sets and piercing multiple nodes simultaneously to graphs without geographical information.
As FlowCutter also achieved quite good results on general graphs albeit with slow running times~\cite{hs-gbpo-18}, this might be an interesting direction for future research.

%%%%%%%%%%%%%%%%%%%%%%%%%%%%%%%%%%%%%%%%%%

%%%%%%%%%%%%%%%%%%%%%%%%%%%%%%%%%%%%%%%%%%
\vspace{6pt} 

%%%%%%%%%%%%%%%%%%%%%%%%%%%%%%%%%%%%%%%%%%
%% optional
%\supplementary{The following are available online at \linksupplementary{s1}, Figure S1: title, Table S1: title, Video S1: title.}


%%%%%%%%%%%%%%%%%%%%%%%%%%%%%%%%%%%%%%%%%%
\funding{This research was partially funded by the DFG under grants WA654/19-2 and WA654/22-2.}
%\funding{Please add: ``This research received no external funding'' or ``This research was funded by NAME OF FUNDER grant number XXX.'' and  and ``The APC was funded by XXX''. Check carefully that the details given are accurate and use the standard spelling of funding agency names at \url{https://search.crossref.org/funding}, any errors may affect your future funding.}

%%%%%%%%%%%%%%%%%%%%%%%%%%%%%%%%%%%%%%%%%%
\acknowledgments{We thank Ben Strasser for helpful discussions and for providing us the setup and code of the experiments conducted in~\cite{hs-gbpo-18}.}%In this section you can acknowledge any support given which is not covered by the author contribution or funding sections. This may include administrative and technical support, or donations in kind (e.g., materials used for experiments).}

%%%%%%%%%%%%%%%%%%%%%%%%%%%%%%%%%%%%%%%%%%
\conflictsofinterest{The authors declare no conflict of interest. The funders had no role in the design of the study; in the collection, analyses, or interpretation of data; in the writing of the manuscript, or in the decision to publish the results.}
%\conflictsofinterest{Declare conflicts of interest or state ``The authors declare no conflict of interest.'' Authors must identify and declare any personal circumstances or interest that may be perceived as inappropriately influencing the representation or interpretation of reported research results. Any role of the funders in the design of the study; in the collection, analyses or interpretation of data; in the writing of the manuscript, or in the decision to publish the results must be declared in this section. If there is no role, please state ``The funders had no role in the design of the study; in the collection, analyses, or interpretation of data; in the writing of the manuscript, or in the decision to publish the results''.}

%%%%%%%%%%%%%%%%%%%%%%%%%%%%%%%%%%%%%%%%%%
%% optional
\appendixtitles{no} %Leave argument "no" if all appendix headings stay EMPTY (then no dot is printed after "Appendix A"). If the appendix sections contain a heading then change the argument to "yes".
%\appendix
%\section{}
%\unskip
%\subsection{}
%The appendix is an optional section that can contain details and data supplemental to the main text. For example, explanations of experimental details that would disrupt the flow of the main text, but nonetheless remain crucial to understanding and reproducing the research shown; figures of replicates for experiments of which representative data is shown in the main text can be added here if brief, or as Supplementary data. Mathematical proofs of results not central to the paper can be added as an appendix.
%
%\section{}
%All appendix sections must be cited in the main text. In the appendixes, Figures, Tables, etc. should be labeled starting with `A', e.g., Figure A1, Figure A2, etc. 

%=====================================
% References, variant B: external bibliography
%=====================================
\reftitle{References}
\externalbibliography{yes}
\bibliography{references}

%%%%%%%%%%%%%%%%%%%%%%%%%%%%%%%%%%%%%%%%%%
%% optional
%\sampleavailability{Samples of the compounds ...... are available from the authors.}

%% for journal Sci
%\reviewreports{\\
%Reviewer 1 comments and authors’ response\\
%Reviewer 2 comments and authors’ response\\
%Reviewer 3 comments and authors’ response
%}

%%%%%%%%%%%%%%%%%%%%%%%%%%%%%%%%%%%%%%%%%%
\end{document}